\documentclass[%
 reprint,
groupedaddress,
showpacs,
 amsmath,amssymb,
 aip,
]{revtex4-1}

\usepackage[final]{graphicx}
\usepackage{dcolumn}
\usepackage{bm}

\begin{document}


\title{Functionalization of carbon nanotubes with -CH$_n$, -NH$_n$ fragments, -COOH and -OH groups }

\author{Karolina Z. Milowska}
 \email{karolina.milowska@gmail.com}
\author{Jacek A. Majewski}%
 \email{Jacek.Majewski@fuw.edu.pl }
\affiliation{%
 Institute of Theoretical Physics, Faculty of Physics, University of Warsaw, ul. Ho\.za 69,
PL-00-681 Warszawa, Poland
\\
}%

\date{\today}

\begin{abstract}
We present results of extensive theoretical studies concerning stability, morphology, and band structure of single wall carbon nanotubes (CNTs) covalently functionalized by -CH$_n$ (for n=2,3,4), -NH$_n$ (for n=1,2,3,4), -COOH and -OH groups. These studies are based on {\sl ab initio} calculations in the framework of the density functional theory. For functionalized systems, we determine the dependence of the binding energies on the concentration of the adsorbed molecules, critical densities of adsorbed molecules, global and local changes in the morphology, and electronic structure paying particular attention to the functionalization induced changes of the band gaps. These studies reveal physical mechanisms that determine stability and electronic structure of functionalized systems and also provide valuable theoretical predictions relevant for application. In particular, we observe that functionalization of CNTs causes generally their elongation and locally sp$^2$ to sp$^3$ rehybridization in the neighborhood of chemisorbed molecules. For adsorbants making particularly strong covalent bonds with the CNTs, such as the -CH$_2$ fragments, we observe formation of the characteristic pentagon/heptagon (5/7) defects. 
In systems functionalized with the -CH$_2$, -NH$_4$, and -OH groups, we determine  critical density of molecules that could be covalently bound to the lateral surface of CNTs. Our studies show that functionalization of CNTs can be utilized for band gap engineering. Functionalization of CNTs can also lead to changes in their metallic/semiconductor character. In semiconducting CNTs, functionalizing molecules such as -CH$_3$, -NH$_2$, -OH, -COOH, and both -OH $\&$ -COOH,  introduce 'impurity' bands in the band gap of pristine CNTs. In the case of -CH$_3$, -NH$_2$ molecules, the induced band gaps are typically smaller than in the pure CNT and depend strongly on the concentration of adsorbants. However, functionalization of semiconducting CNTs with hydroxyl groups leads to the metallization of CNTs. On the other hand, the functionalization of semi-metallic (9,0) CNT with -CH$_2$ molecules causes the increase of the band gap and induces semi-metall to semiconductor transition. 
\end{abstract}

\maketitle

\section{\label{sec:level1}Introduction}

Carbon nanotubes (CNTs) since their discovery in 1991 by Iijima \cite{Iijima1991} have become flagship of the twenty-first century nanoscience, mostly owing to their remarkable electronic, mechanical and thermal properties. Their extraordinary chemical and thermal stability causes that the CNTs are considered as very promising material for the whole plethora of applications. \cite{Terrones2003, book1, book2} One of the highly explored applications of CNTs are nanocomposite materials. Addition of CNTs fibres to various softer  materials such as metallic alloys, polymers, epoxy or cautchouc generally enhances their mechanical strength, electrical and thermal conductivity, chemical stability, absorption of electromagnetic field (for screening), field emission (for field emission displays), and diminishes friction on the composite's surface. \cite{book1, book2, Gou2005, Srivastava2003, Wei2004} The other intensively investigated field of application of CNTs is nanoelectronics, particularly the issue of employing CNTs in very efficient electrical sensors of chemical and/or biological substances. It  is well known already that the sensors based on quantum wires (QWs) are more effective than the standard field effect transistors (FETs) based on heterostructures, owing to the larger surface to volume ratio. The electronic sensors based on CNTs might exhibit even better performance than the traditional semiconductors Si, GaAs, or GaN. 
 
In recent years, significant progress in the applications of CNTs has been achieved. However, there is one important issue that really hampers the larger progress in the above mentioned areas of applications. Pristine CNTs are not soluble in water or in organic solvents and have tendencies to create bundles, mostly because  of dangling $\pi$-bonds of the C atoms on CNT's surface which could build strong covalent bonds among themselves. This behavior leads to non-homogenous distribution of CNTs in the composite's matrix and to certain degree limits the usage of CNTs in nanocomposites on industrial scale. These problems can be cured by the chemical functionalization \cite{Gou2005, Veloso2006, yook2010, carbon2010, Li2004, Rosi2007, Agraval2006, Wongchoosuk2009, Mao1999} of CNTs, i.e., adsorption of chemical substances on the surface, which should lead to homogeneous distribution of CNTs in a composite and additionally should cause stronger bonding of the functionalized CNTs to the composite matrix. It turns out that the CNTs could be successfully doped by many various  groups, fragments and radicals. \cite{Gou2005, yook2010, carbon2010, Kong, Wang, Strano, Stevens, Nakamura}
The functionalization is also important in the context of CNT based sensors. Larger biological systems, like proteins or DNA, do not bind directly to the CNT wall and only functionalization of CNTs with small atomic groups covalently bonded to CNTs promises to  improve significantly intermolecular interactions and hydrophilicity and further to fabricate efficient sensor devices. \cite{Terrones2003, Keren}

The efficient chemical functionalization is a prerequisite for large class of functional materials and devices employing CNTs. However, chemical functionalization also leads to certain unwanted modifications in chemical and physical properties of CNTs. Therefore, the experimental and theoretical studies of functionalized CNTs are crucial for gaining understanding of functionalization processes and can further facilitate the design of novel effective materials and devices. The relevant problems include (i) the stability of functionalized structures, (ii) functionalization induced changes in geometry of the systems, 
(iii) critical density of functionalizing molecules that can be adsorbed at the surface of the CNTs, and (iv) electronic structure of functionalized systems.   

Therefore, in this paper we report results of our extensive theoretical studies on nominally metallic and semiconducting single wall CNTs functionalized with the most popular groups and fragments, -CH$_n$ (for n=2,3,4), -NH$_n$ (for n=1,2,3,4), and -OH, -COOH, at various concentrations of functionalized species. We investigate the stability of the functionalized systems and changes in their morphology and electronic structure. Taking into account very weak experimental knowledge of the systems, our studies could provide a guide for further technological developments. 

To our knowledge, the present work is the first theoretical study of the CNT functionalization issue dealing simultaneously with such broad range of functionalizing molecules and considered properties. However, we are aware of the works providing results for very specific systems. \cite{Veloso2006, stobinski2010, Li2004, Rosi2007, Agraval2006, Doudou, Zhao, Zhao2, Bauschlicher,  Shirvani, appa, aip, diamond} Even if some systems were studied previously, the authors used very often different methodology of their calculations and considered mostly only electronic structure of functionalized CNTs. Our paper constitutes a global description within one single theoretical approach. Therefore, for completeness, we decided to present also results for the systems studied earlier.     

The paper is organized as follows. In Section ~\ref{sec:level2}, we present methodological details of the calculations and provide the description of the studied systems. The results of our calculations are presented in Sec.~\ref{sec:level3}. First, in Sec.~\ref{sec:level3a} we discuss the results of the  present calculations concerning the morphology and stability of functionalized structures with -NH$_n$ fragments (Sec.~\ref{sec:level3aa}), -CH$_n$ hydrocarbons (Sec.~\ref{sec:level3ab}), -OH and -COOH groups (Sec.~\ref{sec:4}), and the lattice constant of functionalized systems in Sec.~\ref{sec:level3ad}. The electronic structure of functionalized groups is discussed in Sec.~\ref{sec:level3b} and finaly the paper is concluded in Sec.~\ref{sec:level4}. 

\section{\label{sec:level2}Calculation details}

As prototypes of the functionalized carbon nanotubes, we study systems consisting of three different types of nominally metallic and semiconducting (9,0), (10,0), (11,0) single wall carbon nanotubes with various concentrations of -CH$_n$ (for n=2,3,4), -NH$_n$ (for n=1,2,3,4) fragments, and -OH, -COOH groups chemisorbed at the side walls of CNTs. 
For hydrocarbyl -CH$_n$, the smallest molecule we consider is -CH$_2$, since there are no experimental evidences that -CH can adsorb to the CNTs. 
Systematically increasing the number of adsorbed molecules, we are able: (i) to find out the critical density of adsorbed molecules, (ii) to observe building of functionalization caused defects and chemically not bound aggregates apart from the CNT surface, and finally (iii) to determine how the properties of the functionalized systems evolve with the density of the functionalizing species. Our studies of the functionalized systems are based on the ground state total energy calculations, which do not include vibrational dynamical contribution to the free energy. However, we have performed very careful search through the configurational space to optimize the geometries of all functionalized systems that guarantee the energetically most favorable configurations.  Comparing stability of various functionalized structures, we always refer to the total energies of these configurations.  

In this paper we use adsorption energy and binding energy as the measure of the stability of the systems studied. The adsorption energy per molecule has been calculated according to the formula
\begin{equation}
E_{ads}  = \frac{1}{N} \left( E_{CNT + groups}  - (E_{CNT}  + N \cdot E_{group} )\right),
\label{eq:1} 
\end{equation}
where $E_{CNT + groups}$ is the total energy of the functionalized system, $E_{CNT}$ the total energy of the pure CNT, N the number of adsorbed groups, and $E_{group}$ the total energy of the free group (i.e., in the vacuum). This quantity can be understood as an average energy required to bind a given group to side surface of CNT. Negative adsorption energy means that the adsorption of groups is energetically favorable and the functionalized system is stable. 
Since we employ a method where Kohn-Sham orbitals are expanded into a set of localized atomic orbitals, we have also calculated basis set superposition error (BSSE) that influences the magnitudes of adsorption energy. We follow typical procedure and compute counterpoise correction \cite{bsse0,bsse1} to adsorption energy according to the formula.
\begin{equation}
E_{cc}  = \frac{1}{N} \left( E_{CNT + \cdot ghost}  - E_{CNT}  + E_{ghost + groups} - N \cdot E_{group} )\right),
\label{eq:bsse} 
\end{equation}
where $E_{CNT +  ghost}$ and $E_{ghost + groups}$ are Kohn Sham energies of the functionalized system but where the adsorbants or nanotube are replaced  by their ghosts \cite{soler}, respectively. For these calculations, we fix the atoms at their optimized positions. It turns out, that $E_{cc}$ might reach up to 44$\%$ of $E_{ads}$ for some adsorbed molecules, however, systematic check through studied adsorbants at various concentrations clearly confirms that all functionalized systems remain stable and the corrected values of the adsorption energies for CNTs functionalized with various molecules follow the trends determined by adsorption energies calculated according to the formula ~\ref{eq:1}. This has been illustrated in Fig.~\ref{fig:fig3}, where the adsorption energies calculated with and without BSSE correction have been plotted for (9,0) CNT functionalized with amines for various concentration of adsorbants.

The binding energy per atom reads
\begin{equation}
E_{bind}  = \frac{1}{N_{a} }\left( {E_{CNT + groups}  - \sum\limits_{\alpha  = 1}^{N_{a}} {E_{atom,\alpha } }  } \right),
\label{eq:2} 
\end{equation}
where $N_a $ is the number of atoms in the system and $E_{atom,\alpha }$ is the total atomic energy of the free atom of type $\alpha$ (C, N, O or H). 

The total energies have been obtained from the {\sl ab initio} calculations in the framework of Kohn-Sham realization of the density functional theory DFT. \cite{Hohenberg1965, kohn} We have used the generalized gradient approximation (GGA) in the form proposed by Perdew \textit{et al.}\cite{Perdew1996} for the exchange correlation density functional, as implemented in the numerical package SIESTA. \cite{ordejon, soler} Since very many of the studied systems contain odd number of electrons in the unit cell, we have systematically used the spin-polarized version of the code in this work. We use the self-consistency mixing rate of 0.1, the convergence criterion for the density matrix of 10$^5$, maximum force tolerance equal to 0.01 eV/\AA, Mesh Cutoff of 300 Ry,  split double zeta basis set with spin polarization, and  1x1x10 k-sampling in Monkhorst Pack scheme.
For the band structure and density of states calculations, the Mesh Cutoff has been increased up to 800 Ry and k-sampling changed to 1x1x33 scheme. For better imaging of density of states, the peak width for broadening the energy eigenvalues has been set to 0.05 eV. 

We use supercell geometry to compute pristine and functionalized CNTs, with the periodicity along the symmetry axis of the pure CNTs. Along the symmetry axis, we choose periodicity of the doubled natural lattice constant resulting from the chiral numbers (n,m) (n = 9, 10, and 11, and m = 0, for the systems studied in this paper). Then the supercells of the pristine (9,0), (10,0), and (11,0) CNTs  contain 72, 80, and 88 carbon atoms, respectively.  To eliminate completely the spurious interaction between neighboring cells, the lateral separation (i.e., lateral lattice constants in the direction perpendicular to the symmetry axis) has been set to 30 \AA, which considerably exceeds doubled diameter of the CNTs, even in the case with adsorbed molecules.  For these pristine CNTs, we have performed calculations to obtain the equilibrium geometry and the ground state energy. 

We have also checked how the size of the supercell and relative orientation of adsorbed molecules influences the values of adsorption energies. For single -NH$_2$ molecule adsorbed to the (9,0) CNT, we have performed calculations employing supercells  equal to the doubled (with 72 C atoms), fourfold (with 144 C atoms), and sixfold (with 216 C atoms) natural unit cells and obtained adsorption energies equal to -0.11 Ry, -0.12 Ry, and -0.12 Ry, respectively. This clearly suggests that even in the case of the smallest supercell the adsorbed molecule can be treated as completely isolated, i.e., not exhibiting a spurious interaction with its images in the neighboring cells. To check the influence of relative orientation of adsorbed molecules, we placed two -OH molecules adsorbed to (9,0) CNT into the supercell with four natural unit cells in such a way, that the two C-O bonds were parallel, orthogonal, and anti-parallel to each other. The adsorbtion energies are equal to -0.168, -0.168, and -0.174 Ry/molecule, respectively, and very weakly depend on the relative orientation of the adsorbands.

\begin{figure}
\includegraphics[width=0.30\textwidth]{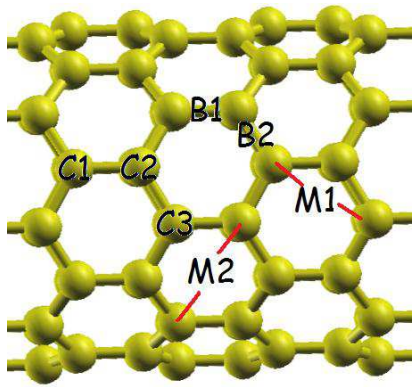}
\caption{\label{fig:fig1}(Color online) Three classes of the sites where adsorbed molecules could be attached to the surface of a CNT: class C - above carbon atoms ($C_1$, $C_2$ and $C_3$); class B - bridge position ($B_1$ and $B_2$); and class M - middle position ($M_1$ and $M_2$).}
\end{figure}

Next, we have performed calculations of functionalized systems with certain number of functionalizing molecules (corresponding to certain concentration) attached to the surface of the CNTs. To minimize interaction between adsorbants, we have placed them as homogeneously and symmetrically as possible around the exterior surface of CNTs. As depicted in Fig. ~\ref{fig:fig1}, there are three classes of sites on CNT's hexagon, indicated as C (C1, C2, and C3), B (B1 $\&$ B2), and M (M1 $\&$ M2) (altogether seven nonequivalent sites) where the functionalizing molecule can be attached to dangling $\pi$-type carbon bond (or bonds) of the nanotube. We have considered all these positions as a starting attachment point for adsorbed molecules in the functionalized systems (placing the adsorbant in such a manner that the distance between the CNT's carbon atoms and the closest atom of the adsorbed molecule was roughly 1.5 \AA), and further we have performed full optimization of the geometry (i.e., the relaxation of atoms around the starting positions) to find out the most energetically stable atomic positions. It turns out that in all cases studied the binding energy after optimization differs by no more than 0.000572 Ry/atom. We observe also the different preferential attachment sites of groups and radicals, namely, the former prefer C site, whereas the later B sites. It is interesting that for none of the chemisorbed molecules the M site is preferable. \cite{Note1}

We would like to note that for pristine CNTs the C1, C2, and C3 sites are equivalent because of the symmetry. However, attached molecules typically do not exhibit the symmetry of the CNT backbone. Therefore, the local optimized geometries of adsorbed molecules to these three sites are not identical. This causes that these three sites are energetically not equivalent when a molecule is attached to them. For example, in the case of -OH molecule attached to C1, C2, and C3 sites of the (9,0) CNTs, the energetically most favorable site is the C2 one. The total energy of the system is by 0.001 eV and 0.13 eV higher in the case of the -OH molecule attached to the C3, and C1 sites, respectively (the internal accuracy of the total energy calculations is better than 0.0001 eV). 
 
For higher concentrations and molecules with larger hydrogen content, we observe also during the geometry optimization process that the fragments of molecules drift outside of the CNT surface. If these fragment are further than 5 \AA $ $   apart from the CNT, they essentially do not influence the functionalized system. 
The details of the geometry of the functionalized systems will be described in the next section.

\section{\label{sec:level3}\textbf{Results and discussion}}

In this section we discuss the stability and the electronic structure of the functionalized carbon nanotubes, presenting the picture that emerges from our calculations and comparing the theoretical results with available experimental data.

\subsection{\label{sec:level3a}Stability and geometry}

We start the presentation of the results by considering the structural properties of the functionalized CNTs. We will discuss the energetics of functionalized systems and the morphology changes induced by the covalent functionalization. We group the results according to the type of functionalizing species and present them in the following order: -NH$_n$ fragments, -CH$_n$ fragments, and -OH $\&$ -COOH groups.

 The results presented in this section correspond to the equilibrium situation at 0 K. They will provide the optimized geometry of the functionalized CNTs, that is basis for calculations of the electronic structure. However, for many cases we have performed molecular dynamics studies (to be published elsewhere) confirming that the functionalized structures remain stable for simulation time of 2 ps and temperatures reaching 1400 K.

\subsubsection{\label{sec:level3aa}\textbf{-NH$_n$} fragments}

\begin{figure} 
\includegraphics[width=0.45\textwidth]{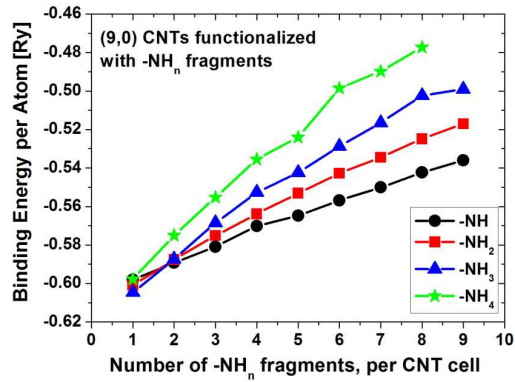}
\caption{\label{fig:fig2} (color online) Binding energies of the functionalized  CNTs 
 for various -NH$_n$ fragments as a function of their density, measured by the number of adsorbants per 72 carbon atoms of a pure (9,0) CNT.}
\end{figure}

\begin{figure} 
\includegraphics[width=0.45\textwidth]{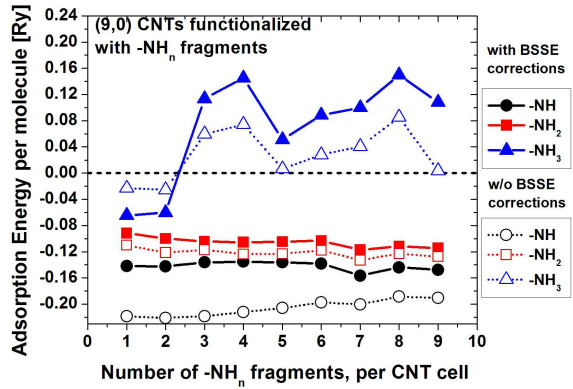}
\caption{\label{fig:fig3} (color online) Adsorption energy per molecule for (9,0) CNTs functionalized with -NH$_n$ fragments as a function of fragments' concentration measured as the number of molecules per double unit cell with 72 atoms. For higher density of NH$_3$ molecules (larger than two molecules per double unit cell), the adsorption energy is positive, just indicating the lack of chemisorption of these fragments to the surface of (9,0) CNT. To illustrate the role of the BSSE correction, we have also plotted the values of the adsorption energies calculated without BSSE correction. These are indicated by empty triangles, squares, and circles for -NH$_3$, -NH$_2$, and -NH fragments, respectively.}
\end{figure}

Amines definitely constitute an important group of the CNTs functionalizing molecules. Mostly owing to their high reactivity, when attached to a CNT,  they play an important role as an anchor of CNT to another material (as in polymer composites), and/or as sticking center for other chemical and biological substances (as in CNT based sensors). \cite{carbon2010, Nakamura, Stevens, Liu99}   Moreover, they induce the change of the hydrophobic nature of pristine CNTs to hydrophilic in the amine functionalized ones, which enables CNTs solubilization.  Functionalization of CNTs with amines could be easily performed using e.g. NH$_3$ plasma treatment of CNTs \cite{yook2010} or substitution of fluorinated single-wall nanotubes (SWNTs) with diamines.\cite{Stevens} It is possible to get even radicals, -NH, bonded to CNTs. Therefore, it is very interesting to determine how the stability of amine functionalized CNTs depends on the concentration of functionalizing -NH$_n$ molecules and whether a critical concentration exists. The main results of our theoretical studies are presented in Figs.~\ref{fig:fig2} and \ref{fig:fig3}, where the binding energy and adsorption energy for (9,0) CNT functionalized with amines are depicted as a function of -NH$_n$ concentration, measured by the number of adsorbed -NH$_n$ molecules per 72 carbon atoms, i.e., doubled unit cell of pristine (9,0) CNT.   

As can be seen in Fig.~\ref{fig:fig2}, the binding energies for (9,0) CNT functionalized with -NH$_n$ molecules for all amines (i.e., -NH, -NH$_2$, NH$_3$, and -NH$_4$) and concentration studied are negative. However, the binding energy generally increases with the number of groups attached to the CNTs (Fig.~\ref{fig:fig2}), which indicates that the binding gets generally weaker with the increasing concentration of adsorbed molecules. This simply indicates that the functionalization of CNT with -NH$_n$ molecules does not lead to the disintegration of the system into free atoms, however, only the adsorption energy as defined in Eq. ~\ref{eq:2} can show whether the -NH$_n$ molecules can be chemisorbed to the CNT wall as a whole species. A glance on Fig.~\ref{fig:fig3} shows that practically only -NH, -NH$_2$ groups are reasonably strongly adsorbed to the CNTs for all studied concentration of adsorbants, with -NH groups being bound more strongly to CNTs than -NH$_2$ ones (-NH radicals exhibit $\&$ lower binding energy than -NH$_2$ ones).  The behavior of NH$_3$ and -NH$_4$ groups is much more complicated and will be discussed later on. 

\begin{figure} 
\includegraphics[width=0.40\textwidth]{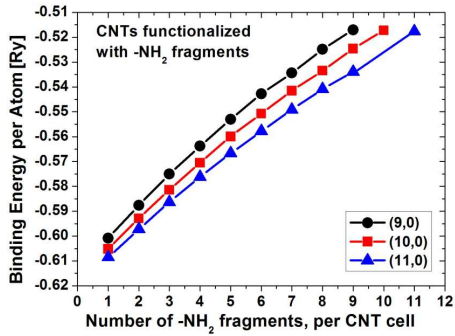}
\caption{\label{fig:fig4} (color online) The binding energy of (9,0), (10,0), and (11,0) CNTs functionalized with -NH$_2$ molecules as a function of the number of -NH$_2$ molecules per 72, 80, and 88 carbon atoms, respectively. Note that the given number of adsorbant molecules constitutes the highest concentration of functionalizing species in the (9,0) CNT and the lowest in the (11,0) CNT.}
\end{figure}

We would like to stress that the described above picture of -NH$_n$ adsorption on the surface of nominally metallic (9,0) CNT remains valid for semiconducting (10,0) and (11,0) CNTs functionalized with -NH$_n$ molecules. As some kind of illustration, we present the binding energy of (9,0), (10,0), and (11,0) CNTs functionalized with -NH$_2$ molecules at various concentrations in Fig. ~\ref{fig:fig4}, where only weak dependence of the binding energy on the CNT type (or diameter) can be observed. Comparison of adsorption energies of all CNTs functionalized with -NH and -NH$_2$ groups reveals that the most stable are (9,0) CNTs, then (10,0), and finally (11,0) CNTs.  Doudou in Ref. \onlinecite{Doudou} shows the same behaviour for zigzag nanotubes functionalized with one -NH$_2$ group.

\begin{figure} 
\includegraphics[width=0.45\textwidth]{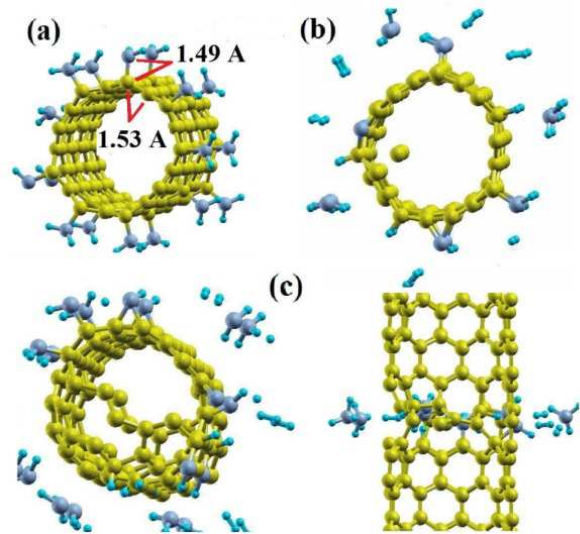}
\caption{\label{fig:fig5} (color online) (a) (9,0) CNT homogeneously decorated by seven -NH$_2$ groups in unit cell. Big yellow spheres indicate carbon atoms, big blue spheres - nitrogen atoms, and small blue spheres hydrogen atoms. The distances between the nearest atoms are also depicted (b) Morphology of (9,0) CNT functionalized with seven -NH$_4$ groups in supercell. Note that C atom from hexagonal ring has been substituted in the hexagonal ring with N atom and has been pushed to the interior of CNT.  (c) Vertical and longitudinal views of CNT decorated with nine -NH$_4$ groups around the CNT. Note that CNT is split into two parts in the place where groups where attached to it, and that leads to creation of the capped nanotubes. }
\end{figure}

Our results clearly show that the functionalization of CNTs with -NH and -NH$_2$ molecules leads to strong chemisorption of the adsorbants. For these systems, the adsorption energy per molecule remains nearly constant as a function of adsorbants concentration (see Fig.~\ref{fig:fig3}), just indicating that CNTs can be functionalized with high density of -NH and -NH$_2$ adsorbants, and we do not find critical density of -NH and -NH$_2$ groups that could be adsorbed on the surface of CNTs without causing destabilization of CNTs. However, -NH$_2$ groups induce small local distortions along the radial direction on the tube sidewall (see Fig.~\ref{fig:fig5}a). These geometry changes are usually described as the local sp$^3$ rehybridization.  Our calculations show that the C-N and C-C bond length are close to the C-C typical distance in the sp$^3$-hybridized diamond (1.54 \AA) and significantly larger than the C-C bond length in the perfect graphene sheet constituting the CNT surface with sp$^2$ hybridization (1.42 \AA), as depicted in Fig.~\ref{fig:fig5}a. Those bond lengths reasonably well compare with other theoretical works,\cite{Doudou, Veloso2006} which give C-N bond length equal to 1.49 \AA).

\begin{figure} 
\includegraphics[width=0.45\textwidth]{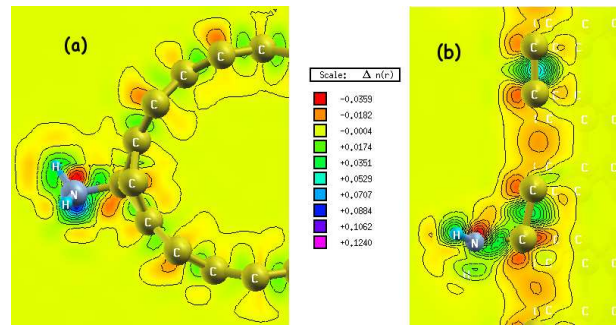}
\caption{\label{fig:fig6} (color online) 
(color online). The difference of the valence pseudocharge density and the superposition of the atomic valence pseudocharge densities for (9,0) CNT functionalized with -NH$_2$ group plotted in the plane perpendicular to the symmetry axis (a) and parallel to the symmetry axis (b), together with the ball-and-stick model of the functionalized system. }
\end{figure}

The strong chemisorption of -NH$_2$ species to CNT surface can be also seen in Fig.~\ref{fig:fig6}, where the distribution of the valence charge is depicted. One can clearly see areas of additional valence electronic charge in the middle of C-N and C-C bonds, and also some excess charge  at N atom. These valence charge rearrangements have mostly local character and the picture does not change essentially with the growing density of the adsorbed molecules.

Let us turn back to the description of CNTs functionalized with NH$_3$ and -NH$_4$ molecules. The chemisorption of NH$_3$ molecules  is possible, strictly speaking, only for two molecules per 72 carbon atoms. For larger number of NH$_3$ molecules the adsorption energy is positive (see Fig.~\ref{fig:fig3}). The fact that the covalent binding of NH$_3$ molecules to CNT surface is not energetically preferable was demonstrated previously in Ref. \onlinecite{Zhao2} and Ref. \onlinecite{Shirvani}.

Concerning the influence of functionalization  with NH$_3$ and -NH$_4$ on the morphology of CNTs, we restrict ourselves to the statement that it is similar to the case of functionalization with the -NH$_2$, i.e., causing mostly local sp$^3$ rehybridization. The most interesting phenomena are observed for CNTs functionalized with -NH$_4$ molecules.  We observe dissociation of -NH$_4$ groups into -NH$_2$ molecules and H$_2$ dimers in the surrounding of the CNT. Further, -NH$_2$ molecules  adsorb to the CNT's surface, whereas H$_2$ molecules remain unbound (see Fig.~\ref{fig:fig5}). The adsorption energy calculated for this case is negative (albeit less negative than in the case of functionalization with -NH$_2$ molecules), however, it has not been depicted in Fig.~\ref{fig:fig3}, since we consider this situation as a special case of  -NH$_2$  functionalization and not the nominal functionalization with -NH$_4$ molecules. For higher concentration of -NH$_4$ groups in the surrounding of (9,0) CNT (starting with around seven groups), we observe characteristic effects (see Fig. ~\ref{fig:fig5}), such as substitution of nitrogen atom into CNT's carbon site and shifting of carbon atom from the hexagonal ring to the inside of the nanotube (Fig.~\ref{fig:fig5}b), or strong deformation of the CNT that leads to splitting it into two parts and capping it simultaneously, as depicted in Fig. ~\ref{fig:fig5}c.  

It has been previously reported that the substitution of nitrogen in zigzag nanotube significantly increases the ionization energy of the tube and can be connected with the electron emitting behavior.\cite{Owens}

In summary of this section, we would like to point out that our calculations clearly demonstrate that metallic and semiconducting nanotubes can be effectively functionalized with -NH and -NH$_2$ molecules, leading to overall stable structures even for relatively high concentration of functionalizing molecules. On the other hand, functionalization with NH$_3$ leads to less stable structures, and functionalization with -NH$_4$ causes large morphological changes of the CNTs. Let us turn now to the studies of CNTs functionalized with -CH$_n$.

\subsubsection{\label{sec:level3ab}\textbf{-CH$_n$} fragments}

Another very important class of molecules that are used for functionalization of CNTs constitute -CH$_n$ groups. Here, we present results of our studies for stability of metallic and semiconducting CNTs functionalized by -CH$_2$ radicals, -CH$_3$ groups, and CH$_4$ molecules. 
We focus on the issue whether there exist critical densities of -CH$_n$  adsorbants that can be chemisorbed on the CNTs' surface. The CNTs have been doped by as many as possible fragments. 

\begin{figure} 
\includegraphics[width=0.40\textwidth]{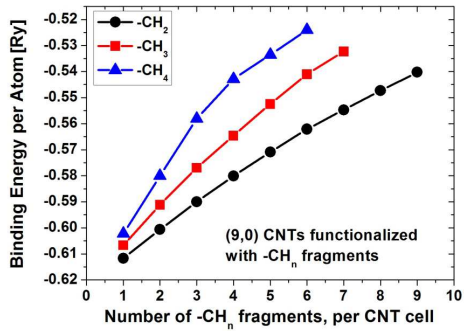}
\caption{\label{fig:fig7} (color online) Binding energies of the functionalized (9,0) CNTs 
 for various -CH$_n$ fragments as a function of their density, measured by the number of adsorbants per 72 carbon atoms of a pure (9,0) CNT.}
\end{figure}

\begin{figure} 
\includegraphics[width=0.40\textwidth]{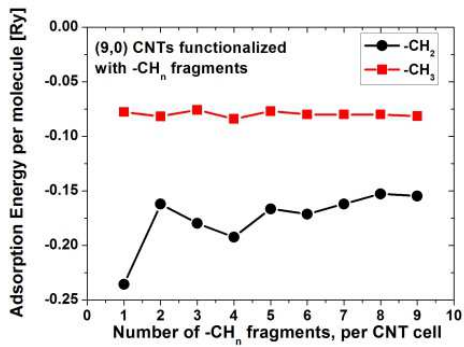}
\caption{\label{fig:fig8} (color online) Adsorption energy per molecule for (9,0) CNTs functionalized with -CH$_2$ and -CH$_3$ fragments.}
\end{figure}

Results of the calculations are presented in Figs.~\ref{fig:fig7} and ~\ref{fig:fig8}, where the binding energy and adsorption energy for (9,0) CNT functionalized with amines are depicted as a function of -CH$_n$ concentration, measured by the number of adsorbed -CH$_n$ molecules per 72 carbon atoms, i.e., doubled unit cell of pristine (9,0) CNT. The binding energy generally increases (i.e., the bonding gets weaker) with the number of -CH$_n$ groups attached to the CNTs (Fig.~\ref{fig:fig7}). The binding energies per atom for -CH$_n$ functionalized (9,0) CNT are very similar to the binding energies for the CNT functionalized with amines (see Fig.~\ref{fig:fig2}). The stability of the functionalized systems is really determined by the adsorption energy depicted in Fig.~\ref{fig:fig8}. It can be seen there that practically only -CH$_2$ fragments are reasonably strongly chemisorbed to the CNT's surface, also at higher concentrations. -CH$_3$ groups bind rather weakly to considered CNTs, and only at moderate densities. CH$_4$ groups as such do not bind at all, but rather dissociates into -CH$_2$ molecule and H$_2$ dimer; where -CH$_2$ chemisorbes at the surface and H$_2$ remains unbound in the surrounding of CNTs (such behavior has been also observed for -NH$_4$ molecules). Zhao \cite{Zhao2} also showed that CH$_4$ do not covalently bind to surface of CNT.
Therefore, we do not include the case of CH$_4$ groups into the figure~\ref{fig:fig8}. 

\begin{figure} 
\includegraphics[width=0.40\textwidth]{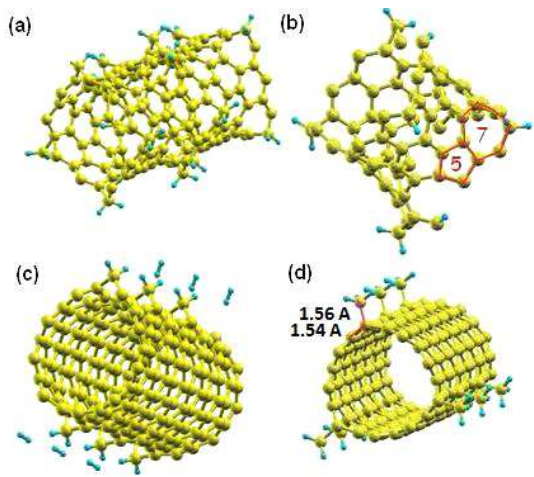}
\caption{\label{fig:fig9} (color online) Morphology of (9,0) CNT functionalized by various -CH$_n$ molecules. Large yellow spheres indicate carbon atoms and small blue ones indicate hydrogen atoms. (a) Geometry of CNT functionalized with -CH$_2$ fragments. Note defected structure of CNT. (b)  Characteristic heptagon/pentagon defects in -CH$_2$ functionalized CNT.  (c) CNT functionalized with CH$_4$ molecules, which dissociate into -CH$_2$ and H$_2$ dimer (d) CNT functionalized with -CH$_3$ group. Note rehybridization from sp$^2$, (graphene) to sp$^3$ (diamond). Length of C-C bond is in \AA. }
\end{figure}

The energetics of the CNTs functionalized with -CH$_n$ is mirrored in the observed geometry of the functionalized systems, as depicted in Fig.~\ref{fig:fig9}. CH$_2$ molecules, which are strongly bound to the (9,0) CNTs, simultaneously cause strong deformation of the CNT's hexagons (see Fig.~\ref{fig:fig9}a) and in some cases introduce characteristic so-called pentagon/heptagon (5/7) defects (see Fig.~\ref{fig:fig9}b). The same effect was reported for the plastic yielding of CNTs.\cite{book1} On the other hand, weakly bound -CH$_3$ molecules have C-C bond lengths of the order of 1.56 \AA (Fig.~\ref{fig:fig9}d). -CH$_3$ groups induce also local distortions along the radial direction on the tube sidewall. It can be understood by the local sp$^3$ rehybridization of the C(from CNT)-C(from group) bonding. Distance between C(CNT)-C(group) bond and C-C bond in hexagonal ring in CNT, are close to the C-C typical distance in the sp$^3$-hybridized diamond (1.54 \AA) and significantly larger than the C-C bond length in the perfect graphene sheet with sp$^2$ hybridization (1.42 \AA). 
Similar results of bond lengths, for armchair CNTs functionalized with -CH$_2$ and -CH$_3$,  were reported in Ref. \onlinecite{Li2004}.

Like in the case of -NH$_4$, the most complicated phenomena occur in the case of functionalization with CH$_4$ molecules. In addition to the previously described dissociation of the CH$_4$ into -CH$_2$ an H$_2$ molecules (see Fig.~\ref{fig:fig9}c), we observe also dissociation of CH$_4$ into weakly bound -CH$_3$ molecule and free hydrogen atom. It indicates the existence of local minima in the configurational space. We have checked it by beginning geometry optimization from various starting positions. However, we have not observed decomposition of CH$_4$ molecule into -CH radical and three hydrogens, reported previously.\cite{Li2004} All these chemical reactions of dissociation are obviously catalyzed by the CNT surface. 
When we place originally CH$_4$ molecules at distances larger than 2 \AA, they do not react with the nanotube surface and do not dissociate. 

\begin{figure} 
\includegraphics[width=0.40\textwidth]{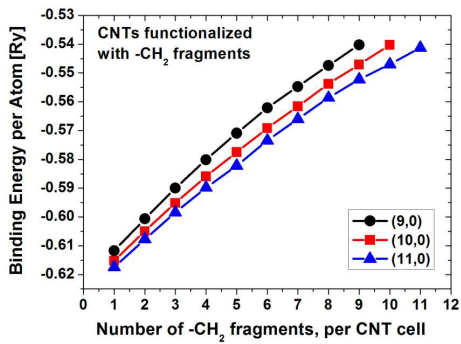}
\caption{\label{fig:fig10} (color online) Binding energy of nominally metallic (9,0), and semiconducting (10,0), and (11,0) CNTs functionalized with -CH$_2$ as a function of the number of adsorbants per 72, 80, and 88 carbon atoms, respectively.}
\end{figure}

Similarly to the case of amines, we have also analyzed the stability and morphology of semiconducting (10,0) and (11,0) CNTs functionalized with -CH$_n$. 
The binding energy the functionalized semiconducting CNTs is similar to the metallic (9,0) CNT functionalized with -CH$_n$ as described above. For an illustration, we plot the binding energy of (9,0), (10,0) and (11,0) CNTs functionalized with -CH$_2$ at various densities in Fig.~\ref{fig:fig10}. Since  (9,0), (10,0) and (11,0) CNTs have also different diameters, one can also conclude that binding energy of -CH$_n$ functionalized CNTs, at least in this range of diameters, is only weakly dependent on the curvature of CNTs. Also the geometrical aspects of (10,0) and (11,0) CNT functionalization with -CH$_3$ and CH$_4$ resemble strongly the case of metallic (9,0) CNT. 

Before we summarize this subsection, we would like to relate our results to other theoretical works. The calculated adsorption energy required to bind -CH$_2$ radical to side surface of (9,0) CNT  is around -0.27 Ry (see Fig.~\ref{fig:fig8} ). This value nicely compares to other theoretical work \cite{Rosi2007} that gives -0.20 Ry. The difference in values results from differences in computational approaches and considered structures. \cite{Note2} In the case of  -CH$_3$ group, we get adsorption energy equal to -0.12 Ry (Rosi in Ref. \onlinecite{Rosi2007} reports the value of -0.03 Ry) for attachment of one group to (9,0) CNT.
Generally, CNT functionalized with -CH$_2$ radicals have more negative adsorption energy then CNT with -CH$_3$ groups. This trend, for armchair nanotubes, was noticed in Ref. \onlinecite{Li2004}, however, in that paper the calculations were performed for quite different [(5,5) and (10,10)] CNTs functionalized with hydrocarbons.

In the summary of this subsection, we would like to stress that the CNTs functionalized with –CH$_2$ radical constitute stable, albeit very often deformed, systems which are reasonable for many applications in which significant densities of adsorbants are reached.

\subsubsection{\label{sec:4} \textbf{-OH} and \textbf{-COOH} groups}

\begin{figure} 
\includegraphics[width=0.45\textwidth]{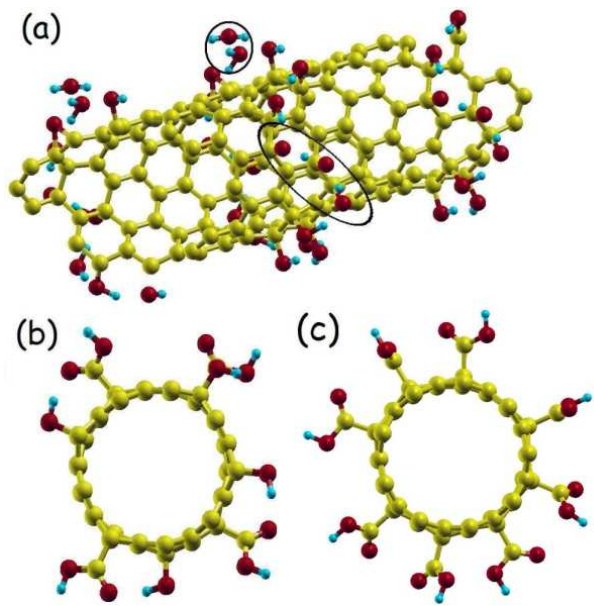}
\caption{\label{fig:fig11} (color online) CNTs functionalized by: (a) -OH groups, (b) -COOH and -OH groups, and (c)-COOH groups. Big yellow spheres indicate carbon atoms, big brown spheres oxygen atoms, and small blue spheres hydrogen atoms. The panel (a) illustrates also formation of the O-H chains (within bigger ellipse), the formation of the water molecules (small ellipse), and the bonding of Oxygen atom in the bridge position (close to the small ellipse).}
\end{figure}

In this subsection, we present results of CNT functionalized by -OH and -COOH groups. Such systems can be easily synthesized \cite{Veloso2006, stobinski2010} and exchanged by other groups using standard chemical reactions allowing one to attach to CNTs more complex molecules like DNA.
We have studied three types of CNT functionalization as illustrated in Fig.~\ref{fig:fig11}: (i) only with -OH groups (Fig.~\ref{fig:fig11}a), (ii) simultaneously with -OH and -COOH groups (Fig.~\ref{fig:fig11}b), and (iii) with -COOH groups (Fig.~\ref{fig:fig11}c). 

\begin{figure} 
\includegraphics[width=0.40\textwidth]{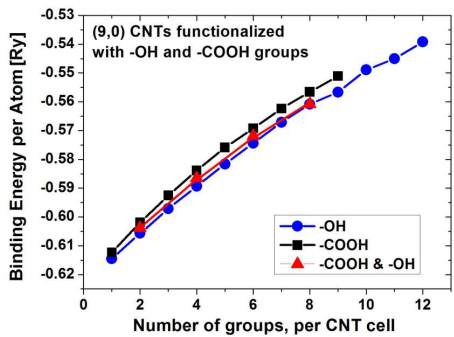}
\caption{\label{fig:fig12} (color online) Binding energy per atom of (9,0) CNT functionalized with (i) -OH molecules (blue dots), (ii) -COOH molecules (black squares), and (iii) simultaneously with -COOH and -OH molecules (red triangles) as a function of number of attached groups.}
\end{figure}

\begin{figure} 
\includegraphics[width=0.45\textwidth]{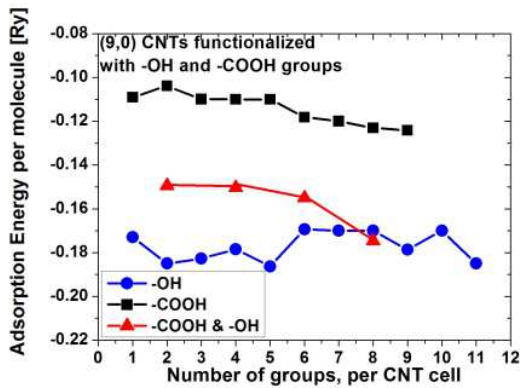}
\caption{\label{fig:fig13} (color online) Adsorption energy per molecule for (9,0) CNT functionalized with (i) -OH molecules (blue dots), (ii) -COOH molecules (black squares), and (iii) simultaneously with -COOH and -OH molecules (red triangles) as a function of number of attached groups.}
\end{figure}

We start presentation of results by demonstrating the dependence of the binding and adsorption energies of (9,0) CNT functionalized in three ways described above as a function of concentration of functionalizing molecules, and depicted respectively in Figs.~\ref{fig:fig12} and \ref{fig:fig13}. As in the previous cases of functionalization discussed so far, the binding energy increases (indicating weaker bonding) with the number of functionalizing molecules attached to the surface of CNT, being for all concentration the lowest for systems functionalized with -OH groups.  

The adsorption energy per molecule follows the trend of the binding energy, being the lowest for the systems functionalized with -OH groups and the weakest for -COOH functionalizing molecules. The value of adsorption energy for one -COOH group bound  to (9,0) CNT (-0.11 Ry, see Fig.~\ref{fig:fig13}) nicely agrees with other theoretical prediction (-0.10 Ry) made by  Zhao.\cite{Zhao} 
 
We have attached up to 18 -OH groups to the 72 atom double unit cell of (9,0) CNT. However, we find out that only 12 -OH groups can be attached uniformly to the CNT. For higher concentrations, owing to steric attraction between -OH groups, the formation of H$_2$O molecules, which are not bounded to the CNT surface, is observed. However, this process does not destroy the structure of the CNT (see Fig.~\ref{fig:fig11}a).
Functionalization with higher concentrations of -OH groups causes also other effects, such as tendency to form O-H net, and oxygen binding in bridge position to two carbon $\pi$ bonds (see Fig.~\ref{fig:fig11}a). 
The calculated distances between O and H atoms lie in the range from 1.17 \AA $ $ to 1.33 \AA, and differ considerably from the distance between O and H atoms in -OH group being equal to ca. 1 \AA. 

We have observed   local sp$^3$ rehybridization of the C(from CNT) - C or O (from group) bonds. C (from CNT)-C(from group) bond lengths are dependent on number of -COOH groups attached to the CNT and vary from 1.55 \AA $ $ to 1.58 \AA. For one carboxyl group attached to (9,0) CNT, we obtain C-C bond length of 1.55 \AA, which excellently agrees with previous theoretical prediction of 1.54 \AA.\cite{Zhao} For the case of -OH groups, C-O bond lengths are shorter and equal 1.45 \AA $ $ independently of the number of adsorbants.
 
\begin{figure} 
\includegraphics[width=0.40\textwidth]{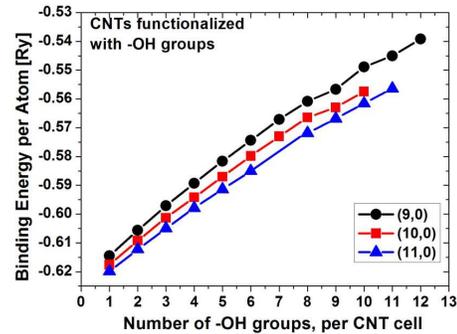}
\caption{\label{fig:fig14} (color online) Binding energy of (9,0), (10,0), and (11,0) CNTs functionalized with -OH molecules as a function of number of -OH adsorbants per 72, 80, and 88 carbon atoms, respectively.}
\end{figure}

The physical picture of functionalization with (i) -OH, (ii) -OH $\&$ -COOH, and (iii) -COOH groups does not change when we consider semiconducting (10,0) and (11,0) CNTs. We exemplify this by showing the binding energy of (9,0), (10,0), and (11,0) CNTs functionalized with -OH molecules as a function of its concentration in Fig.~\ref{fig:fig14}. 

Our calculations demonstrate that the hydroxyl group are very promising functionalizing species leading to stable functionalized CNTs. 

\subsubsection{\label{sec:level3ad}Geometric effects of CNT functionalization}

\begin{figure} 
\includegraphics[width=0.40\textwidth]{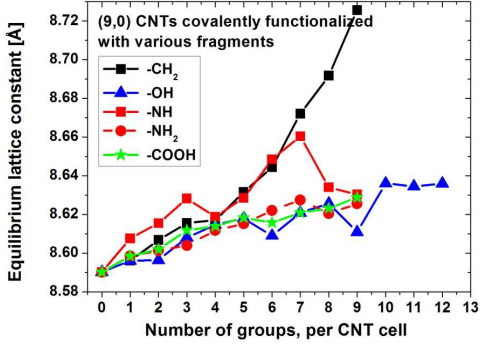}
\caption{\label{fig:fig15} (color online) The equilibrium lattice constant ($l_o $) along symmetry axis (z) of functionalized nanotube as a function of number of attached groups to the sidewall of (9,0) CNTs for -NH, -NH$_2$, -CH$_2$, -OH, and -COOH functionalizing molecules.}
\end{figure}

Before we start discussion of the changes in the electronic structure induced by the functionalization, we would like also to point out one important effect. Namely, functionalization of CNTs not only modifies locally morphology of the functionalized system but also influences the lattice constant (i.e., length) of the functionalized CNTs.
Since the change of the lattice constant alters also the wave vectors, we have investigated this problem in detail. We focus only on molecules that lead to the most stable functionalized structures, namely -NH, -NH$_2$, -CH$_2$, -OH, and -COOH. In Fig.~\ref{fig:fig15} we plot the calculated equilibrium lattice constant ($l_o $) along symmetry axis (z) of the functionalized (9,0) CNT as a function of number of sidewall attached groups. It is clearly seen that the functionalization of (9,0) CNT generally causes elongation of the tube, which increases with the concentration of functionalizing molecules. The most pronounced increase is observed for -CH$_2$ radical. The similar behaviour of the lattice constant is observed for (10,0) and (11,0) functionalized CNTs. 

\subsection{\label{sec:level3b}Electronic structure}

\begin{figure} 
\includegraphics[width=0.40\textwidth]{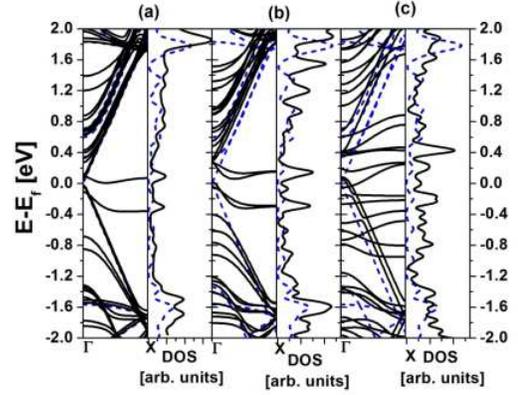}
\caption{\label{fig:fig16} (color online) Electronic band structure along $\Gamma \to X$ line in the Brillouin Zone and the total density of states for (9,0) CNT functionalized with  (a) one, (b) two, and (c) six methyl groups per 72 atom supercell (black solid line). Blue dashed lines indicate the band energies and DOS for the pure (9,0) CNT.  The Fermi energy is set to zero. Note that for all concentrations of adsorbants, the system remains metallic after functionalization.}
\end{figure}

\begin{figure} 
\includegraphics[width=0.40\textwidth]{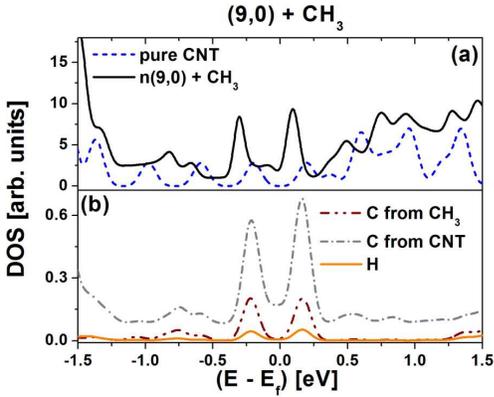}
\caption{\label{fig:fig17} (color online) Band structure for (9,0) CNT functionalized with one -CH$_3$ molecule per 72 atom supercell. (a) - Total DOS  for functionalized (black line) and for pure systems (blue dashed line); (b) - projected DOS on the local orbitals connected to the following atoms:  hydrogen atom from methyl (orange solid line), carbon atom from methyl (brown dotted-dashed line), and carbon atom from the CNT to which methyl molecule is bound (gray dot-dashed line). } 
\end{figure}

\begin{figure} 
\includegraphics[width=0.40\textwidth]{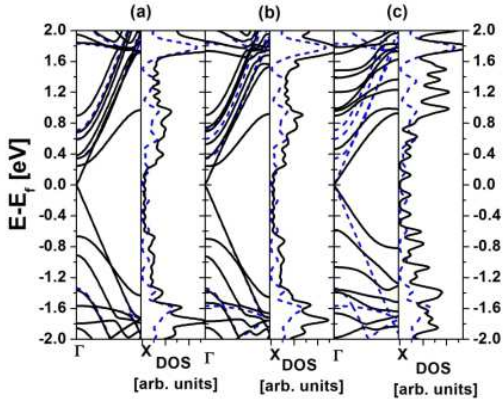}
\caption{\label{fig:fig18} (color online) Electronic band structure along $\Gamma \to X$ line in the Brillouin Zone and the total density of states for (9,0) CNT functionalized with  (a) one, (b) two, and (c) nine -CH$_2$ groups per 72 atom supercell (black solid line). Blue dashed lines indicate the band energies and DOS for the pure (9,0) CNT.  The Fermi energy is set to zero.  Band gap is slightly increased from 0.035 eV for pure (9,0) CNT to 0.05 eV for CNT with three attachments and then drops to 0.04 eV for nine -CH$_2$ radicals. The system remains semi-metallic after functionalization.}
\end{figure}

\begin{figure} 
\includegraphics[width=0.40\textwidth]{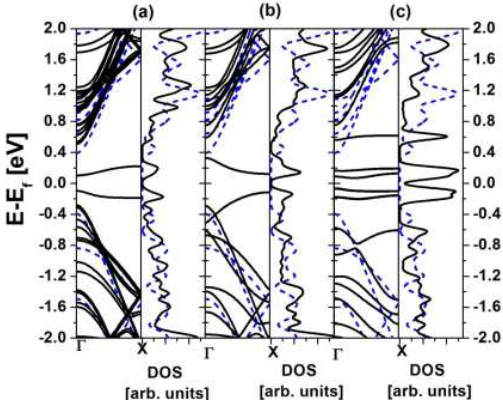}
\caption{\label{fig:fig19} (color online) Electronic band structure along $\Gamma \to X$ line in the Brillouin Zone and the total density of states for (10,0) CNT functionalized with  (a) one, (b) two, and (c) six -CH$_3$ methyl groups per 80 atom supercell (black solid line). Blue dashed lines indicate the band energies and DOS for the pure (10,0) CNT. System remains semiconducting after functionalization, however, the band gap decreases with increasing number of attachments.}
\end{figure}

\begin{figure} 
\includegraphics[width=0.40\textwidth]{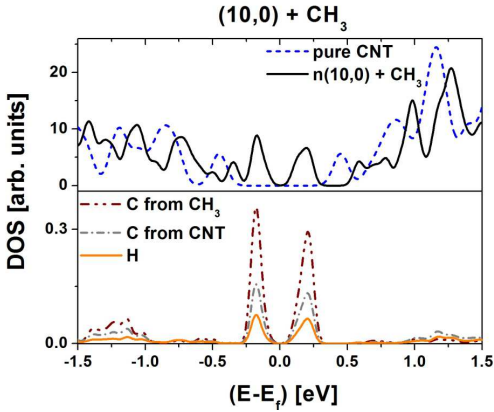}
\caption{\label{fig:fig20} (color online) Band structure for (10,0) CNT functionalized with one -CH$_3$ molecule per 80 atom supercell. (a) - Total DOS of functionalized (black line) and for pure (blue dashed line) (10,0) CNT; (b) - projected DOS on the basis orbitals connected to the following atoms:  hydrogen atom from methyl (orange solid line), carbon atom from methyl (brown dotted-dashed line), and carbon atom from the CNT to which methyl group is bound (gray dot-dashed line). }
\end{figure}

\begin{figure} 
\includegraphics[width=0.40\textwidth]{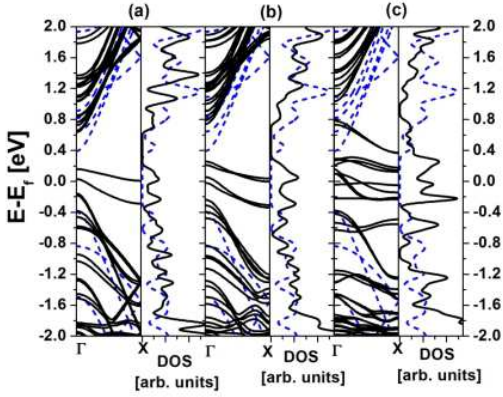}
\caption{\label{fig:fig21} (color online) Electronic band structure along $\Gamma \to X$ line in the Brillouin Zone and the total density of states for (10,0) CNT functionalized with  (a) one, (b) two, and (c) six -OH (hydroxyl) groups per 80 atom supercell (black solid line). Blue dashed lines indicate the band energies and DOS for the pure (10,0) CNT. The Fermi energy is set to zero. Note that all functionalized systems are metallic.}
\end{figure}

\begin{figure} 
\includegraphics[width=0.40\textwidth]{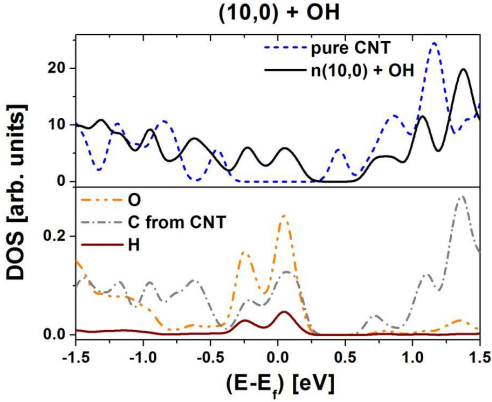}
\caption{\label{fig:fig22} (color online)Band structure for (10,0) CNT functionalized with one -OH group per 80 atom supercell. (a) - Total DOS of functionalized (black line) and for pure (blue dashed line) (10,0) CNT; (b) - projected DOS on the basis orbitals connected to the following atoms:  hydrogen atom from -OH group (orange solid line), oxygen atom from hydroxyl (brown dotted-dashed line), and carbon atom from the CNT to which hydroxyl group is bound (gray dot-dashed line).}
\end{figure}

\begin{figure} 
\includegraphics[width=0.40\textwidth]{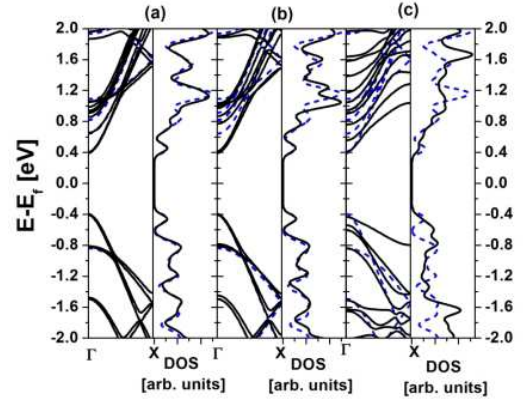}
\caption{\label{fig:fig23} (color online) Band structure and total density of states of (10,0) CNT functionalized (black solid line) with one (a), two (b) and six (c) -CH$_2$ radicals per supercell (80 carbon atom). Blue dashed line applies to pure (10,0) CNT.}
\end{figure}

The covalent functionalization of the CNTs can also lead to changes in the electronic structure. There are very interesting issues: (i) to which extent the band structure of the pristine CNT can be modified by functionalization, and (ii) whether the functionalization procedure can induce a transformation from metallic CNTs into semiconducting ones or vice versa. This aspect of functionalization induced changes of metallic character of CNTs can also have practical meaning. Recently, it has been demonstrated that  covalent functionalization can be used as experimental method \cite{Strano} to separate semiconducting from metallic and semi-metallic (i.e., nominally, on the basis of the chiral numbers $n$ and $m$ metallic, but in reality exhibiting very small energy gap) tubes with high selectivity and scalability. Therefore, in this section, we have decided to present shortly the most characteristic effects of CNT functionalization on the electronic structure. The presented results reveal physical mechanisms and provide valuable quantitative theoretical predictions. 
In the following, the band structure and density of states (DOS) for functionalized systems are always related to 
the band structure and DOS for pristine nominally metallic (or semi-metallic) (9,0) CNTs  and semiconducting (10,0) CNTs. These quantities are plotted on every figure, from Fig.~\ref{fig:fig16} to Fig.~\ref{fig:fig23}, as blue dashed lines.  Our calculations for pure (9,0)  and (10,0) CNTs give band gaps equal to 0.0346 eV  and of 0.7844 eV, respectively. These band gap values compare very well with previous theoretical and experimental works (0.02-0.08 eV for (9,0) \cite{Matsuda, Cao, Kienle, Gulseren} and 0.65-0.88 eV for (10,0) CNT \cite{Doudou,Kienle, Zhao, Zhao2, Bauschlicher}).

We start the presentation of the results by looking at the details of the band structure for the semi-metallic (9,0) CNT functionalized  with methyl (-CH$_3$) group at various concentrations (see Fig.~\ref{fig:fig16}). It is clearly seen that functionalizing molecules considerably modify electronic structure of pure (9,0) CNT. The flat bands in the energetic neighborhood of the band gap of pure tube introduced by adsorbant molecules constitute the most pronounced effect. The number of functionalization induced energy bands is proportional to the concentration of adsorbands.  The Fermi level is set at zero energy and it is easily seen that the functionalized systems remains semi-metallic for all concentration of adsorbands. Since the system (9,0) CNT + methyl group has odd number of electrons, we observe spin-splitting in the electronic band structure. The two 'impurity' bands induced by -CH$_3$ group in the band gap of pure CNT (seen in Fig.~\ref{fig:fig16}a) belong to different spin populations. Here we would like to point out that we do not separate spin majority and spin minority bands in presented plots with the band structure. In the case of two -CH$_3$ groups, one has even number of electrons. However, owing to the slightly nonequivalent position of each -CH$_3$ group on the surface of CNT and some kind of interaction between them, there are four 'impurity' bands in the band gap region of pure (9,0) CNT. 

One can analyze the chemical origin of the 'impurity'  bands by plotting the projected density of states. For the system consisting of (9,0) CNT and -CH$_3$ group, the results of such analysis are depicted in Fig.~\ref{fig:fig17}b.  As one can see the main contribution to 'impurity' bands comes from carbon atom from the tube backbone to which the -CH$_3$ group is bounded. Also the carbon atom from methyl group, and, to a smaller extent, the methyl's hydrogen atoms contribute to the 'impurity' band. Therefore, these 'impurity' states originate from sp$^2$ $\to$ sp$^3$ rehybridization caused by the covalent bonding of the adsorbant to the surface of CNT. This rehybridization has been discussed earlier in the paper. 

We would like to stress that the described physical mechanisms of formation of the 'impurity' bands in the gap of pure (9,0) CNT remain valid for (9,0) CNT functionalized with  -NH$_2$, -OH, -COOH and both -OH $\&$ -COOH. However, the role of the $\pi$ orbital originating from methyl's carbon is now played by nitrogen or oxygen orbitals in the case of -NH$_2$, and hydroxyl and/or carboxyl groups, respectively. 

In the case of the (9,0) CNT functionalizing with -CH$_2$ radical (see Fig.~\ref{fig:fig18}), interestingly one observes the situation that is  completely different from the one described above. There are no impurity states around Fermi level induced by presence of the -CH$_2$, even for the critical concentration of nine groups per 72 carbon atoms. Density of states, as well as band structure are rather similar to pure (9,0) CNT. Our results for those attachments are consistent with earlier studies.\cite{Li2004, Zhao} We have noticed slight change in the band gap magnitude, being for one adsorbed -CH$_2$ group slightly lower than in the case of pure CNT  (0.01 eV  vs. 0.035 eV for pure CNT) and a little bit bigger for higher concentrations of these fragments (it is equal to 0.05 eV for three CH$_2$ molecules and saturates at the value of 0.04 eV for higher number of adsorbants). Also appearance of the so-called 5/7 defects (see Sec.~\ref{sec:level3ab}) in the functionalized system alters the band structure only moderately and do not introduce 'impurity' bands within the band gap of pristine (9,0) CNT. Further, the analysis of the projected DOS indicates that the states induced by those functionalizing molecules lay deeply under Fermi level.

As we have already indicated,  the band gap of (9,0) CNT functionalized with -CH$_2$  reaches the maximum for $N$ = 3 (i.e., for three functionalizing molecules).  For  $N>$  3,  we observe that the band gap minimally decreases with increasing number of radicals attached to the surface of CNT. As it was shown in Sec. ~\ref{sec:level3ab}, -CH$_2$ radicals cause big deformations of lateral surface of functionalized nanotube. Therefore, this situation could be comparable with behavior of  stressed (9,0) CNT,\cite{Gulseren} where the band gap starts to decrease with increasing strain. It was shown by Gulseren \cite{Gulseren} and Silva \cite{Silva} that applying  uniaxial or torsional strain to SWNT can lead to change of metallic/semiconducting character of CNT. Moreover, theoretically \cite{Gulseren, Kienle} and experimentally \cite{Strano} it has been observed that stress can cause charge reorganization with radial deformation which directly modifies chemical reactivity of the surface of CNT. Functionalization which brings out global changes in structure is confirmed by Strano \cite{Strano} to induce such enhancement in reactivity of CNTs. It is understandable in terms of  availability of electrons near the Fermi level which are required to stabilize charge-transfer transitions preceding bond formation. 

For the case of (9,0) and (10,0) functionalized with -NH radical, we observe oscillating magnitude of  the band gap with the number of attached molecules. This effect is qualitatively similar the the effect described above for CNT functionalized with -CH$_2$. Semi-metallic (9,0) CNT with small concentration of -NH has smaller band gap than the pure (9,0) CNT and can still be considered as semi-metallic system. 

Having described the influence of covalent functionalization on semi-metallic CNT, we turn now towards functionalization induced band structure changes in semiconducting CNTs. As a prototype of semiconducting nanotube we consider the (10,0) SWNT. As in the case of semi-metallic (9,0) CNT, we analyze mostly two groups of functionalizing molecules, the first group introduces the changes in the electronic structure in the energy range close to the Fermi energy (e.g., -CH$_3$, -NH$_2$, -OH, and -COOH) and the second group of adsorbands that rather modify only the density of valence and conduction bands (e.g., -CH$_2$).  

Let us look at the details of the band structure for the semiconducting (10,0) CNT functionalized  with methyl (-CH$_3$) group at various concentrations (see Fig.~\ref{fig:fig19}). It is clearly seen that functionalizing molecules considerably modify electronic structure of pure (10,0) CNT. The flat bands introduced by adsorbant molecules in the band gap of pure tube  constitute the most pronounced effect. The number of functionalization induced energy bands is proportional to the concentration of adsorbands.  The Fermi level is set at zero energy and it is easily seen that the functionalized systems remains semiconducting for all concentration of adsorbands. However, in the functionalized (10,0) CNT the highest occupied and the lowest unoccupied band are induced by states of the functionalizing molecules. As a result of growing interaction between bonding and anti-bonding states of adsorbed molecules, the energy gap decreases now with the concentration of adsorbants (see Fig. ~\ref{fig:fig19}). On the other hand, the CNTs with higher concentration of  -CH$_3$ groups have positive adsorption energies, so one should  not expect the metallization of semiconducting (10,0) CNT by high degree of functionalization. 

One can analyze the chemical origin of the 'impurity'  bands by plotting the projected density of states. For the system consisting of (10,0) CNT and -CH$_3$ group, the results of such analysis are depicted in Fig.~\ref{fig:fig20}b.  As one can see the main contribution to 'impurity' bands comes from carbon atom from the tube backbone to which the -CH$_3$ group is bounded. Also the carbon atom from methyl group, and, to a smaller extent, the methyl's hydrogen atoms contribute to the 'impurity' band. Therefore, these 'impurity' states originate from sp$^2$ $\to$ sp$^3$ rehybridization caused by the covalent bonding of the adsorbant to the surface of CNT. The analysis of the chemical nature of the 'impurity' bands clearly shows that for semiconducting (10,0) CNTs and semi-metallic (9,0) ones the chemical nature of the 'impurity' bands is very similar. 

The physical mechanisms that cause the band structure changes in the semiconducting (10,0) CNTs functionalized with -NH$_2$, -OH, -COOH, and -OH $\&$ -COOH are quite similar to the just described above. It is illustrated for the -OH group in Figs.~\ref{fig:fig21} and ~\ref{fig:fig22}.
The (10,0) CNTs functionalized with -NH$_2$ groups exhibit nearly identical behavior as (10,0) CNTs functionalized with methyl molecules concerning the electronic structure. The band gap decreases, however, we do not observe metallization, at least for the concentrations studied. On the other hand, the systems functionalized with -OH (see Fig.~\ref{fig:fig21}), -COOH, and with both -OH $\&$ -COOH exhibit metallic character for all concentrations. In Fig.~\ref{fig:fig22}, it is seen that the metallicity originates from the overlap of HOMO and LUMO induced 'impurity' bands. These results clearly show which type of molecules should be used in order to make semiconducting nanotubes metallic and give important practical hint for technological applications.

The changes of the band structure of semiconducting CNTs functionalized with -CH$_2$ radicals are qualitatively very similar to the case of metallic CNTs functionalized with these species. As an illustration, we show the band structure of (10,0) CNT functionalized with -CH$_2$ molecules at various concentrations in Fig.~\ref{fig:fig23}. These adsorbands do not induce any impurity bands around Fermi level, but only modify the DOS of valence and conduction bands. Band gap of functionalized systems is very slightly larger than in the pure (10,0) CNT and oscillates around 0.82 eV (for up to 6 radicals per supercell shown in Fig.~\ref{fig:fig23}). However, for higher concentration of adsorbands, we observe that the band gap decreases, specifically, for nine -CH$_2$ radicals per 80 carbon atoms, the band gap is reduced to 0.47 eV.

The results of our studies shed light on physical mechanisms leading to the changes of the band structure in  covalently functionalized metallic and semiconducting CNTs and provide also quantitative theoretical predictions for the band gaps in CNTs functionalized with wide range of molecules which are of importance for many CNT applications. The importance of the problem lead many authors to deal with this topic. Astonishingly, there are more papers dealing with the band structure of functionalized CNT than with their stability. Here we would like to bring our results in the context of previous calculations.  
Many of the systems studied have odd number of electrons in the chosen unit cell.
In  contrary to other papers dealing with this issue, \cite{Zhao, Veloso2006, Zhao2, Li2004, Doudou} we have performed fully spin-polarized calculation for all systems studied (with odd and even number of electrons). The resulting band structure for systems with odd number od electrons exhibits typical spin splitting with one band for majority spin (say spin up) completely occupied and one band with minority spin (say spin down)  completely empty. Neglect of spin polarization for a sytem with odd number of electrons (as done in previous studies) leads to one half-filled band.    
To facilitate comparison with other authors, we have performed calculations with and without spin-polarization for the same systems which were analysed by  Veloso \cite{Veloso2006} (two units of (8,0) per supercell functionalized  with one -NH$_2$) and Doudou \cite{Doudou} (three units of (10,0) functionalized with one -NH$_2$). Disabling spin-polarization results in one half-filled impurity band crossed by the Fermi level and leads to the results very similar to the results of Ref. \onlinecite{Veloso2006} and Ref. \onlinecite{Doudou}. 

Electronic structure of CNTs functionalized with molecules such as CH$_4$ or NH$_3$ has been discussed previously.\cite{Bauschlicher,Shirvani, Zhao, Agraval2006} Since our studies of stability of such systems clearly indicate that these systems are not covalently bond to lateral surface of CNT, we have postponed the discussion of their band structure in this paper. Our studies show that the week coupling that between $\pi$ electrons from tube and adsorbands is not strong enough to change considerably DOS or band structure of these systems. 

\section{\label{sec:level4}Conclusions}

We have performed extensive {\sl ab initio} studies of the stability and electronic structure of metallic and semiconducting CNTs functionalized with various molecular groups. We present results for three (9,0), (10,0), and (11,0) single wall carbon nanotubes functionalized with -OH, -COOH, -CH$_n$, and -NH$_n$ molecules, which have been homogeneously distributed on the sidewall of CNTs. These results shed light on physical mechanisms which determine the stability and electronic band structure of functionalized systems and provide valuable theoretical predictions of considerable importance for practical applications. 

The functionalization of semi-metallic and semiconducting CNTs results in global and local modifications of their geometry. Generally functionalized CNTs are elongated and they exhibit covalent bonding between CNT and adsorbants, manifesting itself in local rehybridization and deformation of the functionalized system. The smallest changes in structure of CNTs are caused by attaching -OH and -COOH groups, the strongest by attaching radicals. We observe characteristic pentagon/heptagon (5/7) defects \cite{book1, Thostenson2001} in CNTs functionalized by -CH$_2$ fragments and effect of some kind of 'capping' (or closing) of the nanotube by sufficiently high concentration of  -NH$_4$ molecules. We have determined the critical density of the -CH$_2$, -NH$_4$ fragments, and -OH groups which could be bound to SWNT. 

Functionalization of CNTs changes their electronic structure and allows for band gap engineering. Most of the considered adsorbands: -OH, -COOH, -NH$_2$, and -CH$_3$ introduce impurity bands around Fermi level which originate from states of adsorbed molecules and carbon tube atom to which the molecule is attached.  Functionalization can lead to changes of metallic/semiconductor character of CNTs.  Specifically, semiconducting (10,0) CNT can be transfered into metallic one by functionalization with -OH, -COOH, and/or both -OH $\&$ -COOH, and also for big concentration of -NH$_2$ groups. On contrary, functionalization with -CH$_2$ opens slightly the band gap in semi-metallic (9,0) CNT, just inducing transition to semiconducting CNT. In the systems with odd number of electrons, we have observed half-metallic bahavior, with the valence top band occupied by majority spin electrons and the first empty band having minority spin character. 

We are aware that the functionalized systems could exhibit some degree of disorder that is not taken into account in our supercell based calculations. However, in the situation of nearly complete lack of experimental data, the theoretical predictions presented in this paper provide physical understanding of the functionalization processes and should facilitate the design of functionalized CNT for purpose of application either in composite materials or nanoelectronics.   

\section{\label{sec:level5}Acknowledgments}

The authors acknowledge financial support of the SiCMAT Project financed under the European Founds for Regional Development (Contract No. UDA-POIG.01.03.01-14-155/09). In the very early stages of this work, the financial support was provided by the Polish Council for Science through the Development Grants for the years 2008-2011 (Grant Nos. 15-0011-04/2008 and KB/72/13447/IT1-B/U/08). We thank also PL-Grid Infrastructure and Interdisciplinary Centre for Mathematical and Computational Modeling of University of Warsaw (Grant No. G47-5) for providing computer facilities.


\begin{thebibliography}{45}%
\makeatletter
\providecommand \@ifxundefined [1]{%
 \@ifx{#1\undefined}
}%
\providecommand \@ifnum [1]{%
 \ifnum #1\expandafter \@firstoftwo
 \else \expandafter \@secondoftwo
 \fi
}%
\providecommand \@ifx [1]{%
 \ifx #1\expandafter \@firstoftwo
 \else \expandafter \@secondoftwo
 \fi
}%
\providecommand \natexlab [1]{#1}%
\providecommand \enquote  [1]{``#1''}%
\providecommand \bibnamefont  [1]{#1}%
\providecommand \bibfnamefont [1]{#1}%
\providecommand \citenamefont [1]{#1}%
\providecommand \href@noop [0]{\@secondoftwo}%
\providecommand \href [0]{\begingroup \@sanitize@url \@href}%
\providecommand \@href[1]{\@@startlink{#1}\@@href}%
\providecommand \@@href[1]{\endgroup#1\@@endlink}%
\providecommand \@sanitize@url [0]{\catcode `\\12\catcode `\$12\catcode
  `\&12\catcode `\#12\catcode `\^12\catcode `\_12\catcode `\%12\relax}%
\providecommand \@@startlink[1]{}%
\providecommand \@@endlink[0]{}%
\providecommand \url  [0]{\begingroup\@sanitize@url \@url }%
\providecommand \@url [1]{\endgroup\@href {#1}{\urlprefix }}%
\providecommand \urlprefix  [0]{URL }%
\providecommand \Eprint [0]{\href }%
\providecommand \doibase [0]{http://dx.doi.org/}%
\providecommand \selectlanguage [0]{\@gobble}%
\providecommand \bibinfo  [0]{\@secondoftwo}%
\providecommand \bibfield  [0]{\@secondoftwo}%
\providecommand \translation [1]{[#1]}%
\providecommand \BibitemOpen [0]{}%
\providecommand \bibitemStop [0]{}%
\providecommand \bibitemNoStop [0]{.\EOS\space}%
\providecommand \EOS [0]{\spacefactor3000\relax}%
\providecommand \BibitemShut  [1]{\csname bibitem#1\endcsname}%
\let\auto@bib@innerbib\@empty
\bibitem [{\citenamefont {Iijima}(1991)}]{Iijima1991}%
  \BibitemOpen
  \bibfield  {author} {\bibinfo {author} {\bibfnamefont {S.}~\bibnamefont
  {Iijima}},\ }\href@noop {} {\bibfield  {journal} {\bibinfo  {journal}
  {Nature}\ }\textbf {\bibinfo {volume} {354}},\ \bibinfo {pages} {56}
  (\bibinfo {year} {1991})}\BibitemShut {NoStop}%
\bibitem [{\citenamefont {Terrones}(2003)}]{Terrones2003}%
  \BibitemOpen
  \bibfield  {author} {\bibinfo {author} {\bibfnamefont {M.}~\bibnamefont
  {Terrones}},\ }\href@noop {} {\bibfield  {journal} {\bibinfo  {journal}
  {Annu. Rev. Mater. Res.}\ }\textbf {\bibinfo {volume} {33}},\ \bibinfo
  {pages} {419} (\bibinfo {year} {2003})}\BibitemShut {NoStop}%
\bibitem [{\citenamefont {Goddard}\ \emph {et~al.}(2003)\citenamefont
  {Goddard}, \citenamefont {Brenner}, \citenamefont {Lyshevski},\ and\
  \citenamefont {Iafrate}}]{book1}%
  \BibitemOpen
  \bibfield  {author} {\bibinfo {author} {\bibfnamefont {W.~A.}\ \bibnamefont
  {Goddard}}, \bibinfo {author} {\bibfnamefont {D.~W.}\ \bibnamefont
  {Brenner}}, \bibinfo {author} {\bibfnamefont {S.~E.}\ \bibnamefont
  {Lyshevski}}, \ and\ \bibinfo {author} {\bibfnamefont {G.~J.}\ \bibnamefont
  {Iafrate}},\ }\href@noop {} {\emph {\bibinfo {title} {Handbook of
  Nanoscience, Engineering, and Technology (Electrical Engineering
  Handbook)}}}\ (\bibinfo  {publisher} {CRC Press LLC},\ \bibinfo {year}
  {2003})\BibitemShut {NoStop}%
\bibitem [{\citenamefont {Mitin}\ \emph {et~al.}(2008)\citenamefont {Mitin},
  \citenamefont {Kochelap},\ and\ \citenamefont {Stroscio}}]{book2}%
  \BibitemOpen
  \bibfield  {author} {\bibinfo {author} {\bibfnamefont {V.~V.}\ \bibnamefont
  {Mitin}}, \bibinfo {author} {\bibfnamefont {V.~A.}\ \bibnamefont {Kochelap}},
  \ and\ \bibinfo {author} {\bibfnamefont {M.~A.}\ \bibnamefont {Stroscio}},\
  }\href@noop {} {\emph {\bibinfo {title} {Introduction to nanoelectronic}}}\
  (\bibinfo  {publisher} {Cambridge University Press},\ \bibinfo {year}
  {2008})\BibitemShut {NoStop}%
\bibitem [{\citenamefont {Gou}\ \emph {et~al.}(2005)\citenamefont {Gou},
  \citenamefont {Liangb}, \citenamefont {Zhangb},\ and\ \citenamefont
  {Wang}}]{Gou2005}%
  \BibitemOpen
  \bibfield  {author} {\bibinfo {author} {\bibfnamefont {J.}~\bibnamefont
  {Gou}}, \bibinfo {author} {\bibfnamefont {Z.}~\bibnamefont {Liangb}},
  \bibinfo {author} {\bibfnamefont {C.}~\bibnamefont {Zhangb}}, \ and\ \bibinfo
  {author} {\bibfnamefont {B.}~\bibnamefont {Wang}},\ }\href@noop {} {\bibfield
   {journal} {\bibinfo  {journal} {Composites: Part B}\ }\textbf {\bibinfo
  {volume} {36}},\ \bibinfo {pages} {524} (\bibinfo {year} {2005})}\BibitemShut
  {NoStop}%
\bibitem [{\citenamefont {Srivastava}\ \emph {et~al.}(2003)\citenamefont
  {Srivastava}, \citenamefont {Wei},\ and\ \citenamefont
  {Cho}}]{Srivastava2003}%
  \BibitemOpen
  \bibfield  {author} {\bibinfo {author} {\bibfnamefont {D.}~\bibnamefont
  {Srivastava}}, \bibinfo {author} {\bibfnamefont {C.}~\bibnamefont {Wei}}, \
  and\ \bibinfo {author} {\bibfnamefont {K.}~\bibnamefont {Cho}},\ }\href@noop
  {} {\bibfield  {journal} {\bibinfo  {journal} {Appl. Mech. Rev.}\ }\textbf
  {\bibinfo {volume} {56}},\ \bibinfo {pages} {215} (\bibinfo {year}
  {2003})}\BibitemShut {NoStop}%
\bibitem [{\citenamefont {Wei}\ and\ \citenamefont
  {Srivastava}(2004)}]{Wei2004}%
  \BibitemOpen
  \bibfield  {author} {\bibinfo {author} {\bibfnamefont {C.}~\bibnamefont
  {Wei}}\ and\ \bibinfo {author} {\bibfnamefont {D.}~\bibnamefont
  {Srivastava}},\ }\href@noop {} {\bibfield  {journal} {\bibinfo  {journal}
  {Appl. Phys. Lett.}\ }\textbf {\bibinfo {volume} {85}},\ \bibinfo {pages}
  {2208} (\bibinfo {year} {2004})}\BibitemShut {NoStop}%
\bibitem [{\citenamefont {Veloso}\ \emph {et~al.}(2006)\citenamefont {Veloso},
  \citenamefont {Filho}, \citenamefont {Filho}, \citenamefont {Fagan},\ and\
  \citenamefont {Mota}}]{Veloso2006}%
  \BibitemOpen
  \bibfield  {author} {\bibinfo {author} {\bibfnamefont {M.~V.}\ \bibnamefont
  {Veloso}}, \bibinfo {author} {\bibfnamefont {A.~G.~S.}\ \bibnamefont
  {Filho}}, \bibinfo {author} {\bibfnamefont {J.~M.}\ \bibnamefont {Filho}},
  \bibinfo {author} {\bibfnamefont {S.~B.}\ \bibnamefont {Fagan}}, \ and\
  \bibinfo {author} {\bibfnamefont {R.}~\bibnamefont {Mota}},\ }\href@noop {}
  {\bibfield  {journal} {\bibinfo  {journal} {Chem. Phys. Lett.}\ }\textbf
  {\bibinfo {volume} {430}},\ \bibinfo {pages} {71} (\bibinfo {year}
  {2006})}\BibitemShut {NoStop}%
\bibitem [{\citenamefont {Yook}\ \emph {et~al.}(2010)\citenamefont {Yook},
  \citenamefont {Jun},\ and\ \citenamefont {Kwak}}]{yook2010}%
  \BibitemOpen
  \bibfield  {author} {\bibinfo {author} {\bibfnamefont {J.~Y.}\ \bibnamefont
  {Yook}}, \bibinfo {author} {\bibfnamefont {J.}~\bibnamefont {Jun}}, \ and\
  \bibinfo {author} {\bibfnamefont {S.}~\bibnamefont {Kwak}},\ }\href@noop {}
  {\bibfield  {journal} {\bibinfo  {journal} {App. Surface Science}\ }\textbf
  {\bibinfo {volume} {256}},\ \bibinfo {pages} {6941} (\bibinfo {year}
  {2010})}\BibitemShut {NoStop}%
\bibitem [{\citenamefont {Chidawanyika}\ and\ \citenamefont
  {Nyokong}(2010)}]{carbon2010}%
  \BibitemOpen
  \bibfield  {author} {\bibinfo {author} {\bibfnamefont {W.}~\bibnamefont
  {Chidawanyika}}\ and\ \bibinfo {author} {\bibfnamefont {T.}~\bibnamefont
  {Nyokong}},\ }\href@noop {} {\bibfield  {journal} {\bibinfo  {journal}
  {Carbon}\ }\textbf {\bibinfo {volume} {48}},\ \bibinfo {pages} {2831}
  (\bibinfo {year} {2010})}\BibitemShut {NoStop}%
\bibitem [{\citenamefont {Li}\ \emph {et~al.}(2004)\citenamefont {Li},
  \citenamefont {Y.~Xia}, \citenamefont {X.~Liu},\ and\ \citenamefont
  {Z.~Tan}}]{Li2004}%
  \BibitemOpen
  \bibfield  {author} {\bibinfo {author} {\bibfnamefont {F.}~\bibnamefont
  {Li}}, \bibinfo {author} {\bibfnamefont {M.-Z.}\ \bibnamefont {Y.~Xia}},
  \bibinfo {author} {\bibfnamefont {B.~H.}\ \bibnamefont {X.~Liu}}, \ and\
  \bibinfo {author} {\bibfnamefont {Y.~J.}\ \bibnamefont {Z.~Tan}},\
  }\href@noop {} {\bibfield  {journal} {\bibinfo  {journal} {Phys. Rev. B}\
  }\textbf {\bibinfo {volume} {69}},\ \bibinfo {pages} {165415} (\bibinfo
  {year} {2004})}\BibitemShut {NoStop}%
\bibitem [{\citenamefont {Rosi}\ and\ \citenamefont {Jr.}(2007)}]{Rosi2007}%
  \BibitemOpen
  \bibfield  {author} {\bibinfo {author} {\bibfnamefont {M.}~\bibnamefont
  {Rosi}}\ and\ \bibinfo {author} {\bibfnamefont {C.~W.~B.}\ \bibnamefont
  {Jr.}},\ }\href@noop {} {\bibfield  {journal} {\bibinfo  {journal} {Chem.
  Phys. Lett.}\ }\textbf {\bibinfo {volume} {437}},\ \bibinfo {pages} {99}
  (\bibinfo {year} {2007})}\BibitemShut {NoStop}%
\bibitem [{\citenamefont {Agraval}\ \emph {et~al.}(2006)\citenamefont
  {Agraval}, \citenamefont {Agrawal}, \citenamefont {Singh},\ and\
  \citenamefont {Srivastava}}]{Agraval2006}%
  \BibitemOpen
  \bibfield  {author} {\bibinfo {author} {\bibfnamefont {B.~K.}\ \bibnamefont
  {Agraval}}, \bibinfo {author} {\bibfnamefont {S.}~\bibnamefont {Agrawal}},
  \bibinfo {author} {\bibfnamefont {S.}~\bibnamefont {Singh}}, \ and\ \bibinfo
  {author} {\bibfnamefont {R.}~\bibnamefont {Srivastava}},\ }\href@noop {}
  {\bibfield  {journal} {\bibinfo  {journal} {J. Phys.: Condens. Matter.}\
  }\textbf {\bibinfo {volume} {18}},\ \bibinfo {pages} {4649} (\bibinfo {year}
  {2006})}\BibitemShut {NoStop}%
\bibitem [{\citenamefont {Wongchoosuk}\ \emph {et~al.}(2009)\citenamefont
  {Wongchoosuk}, \citenamefont {Udomvech},\ and\ \citenamefont
  {Kerdcharoen}}]{Wongchoosuk2009}%
  \BibitemOpen
  \bibfield  {author} {\bibinfo {author} {\bibfnamefont {C.}~\bibnamefont
  {Wongchoosuk}}, \bibinfo {author} {\bibfnamefont {A.}~\bibnamefont
  {Udomvech}}, \ and\ \bibinfo {author} {\bibfnamefont {T.}~\bibnamefont
  {Kerdcharoen}},\ }\href@noop {} {\bibfield  {journal} {\bibinfo  {journal}
  {Current. Appl. Phys.}\ }\textbf {\bibinfo {volume} {9}},\ \bibinfo {pages}
  {352} (\bibinfo {year} {2009})}\BibitemShut {NoStop}%
\bibitem [{\citenamefont {Mao}\ \emph {et~al.}(1999)\citenamefont {Mao},
  \citenamefont {Garg},\ and\ \citenamefont {Sinnott}}]{Mao1999}%
  \BibitemOpen
  \bibfield  {author} {\bibinfo {author} {\bibfnamefont {Z.}~\bibnamefont
  {Mao}}, \bibinfo {author} {\bibfnamefont {A.}~\bibnamefont {Garg}}, \ and\
  \bibinfo {author} {\bibfnamefont {S.~B.}\ \bibnamefont {Sinnott}},\
  }\href@noop {} {\bibfield  {journal} {\bibinfo  {journal} {Nanotechology}\
  }\textbf {\bibinfo {volume} {10}},\ \bibinfo {pages} {273} (\bibinfo {year}
  {1999})}\BibitemShut {NoStop}%
\bibitem [{\citenamefont {Kong}\ \emph {et~al.}(2001)\citenamefont {Kong},
  \citenamefont {Chapline},\ and\ \citenamefont {Dai}}]{Kong}%
  \BibitemOpen
  \bibfield  {author} {\bibinfo {author} {\bibfnamefont {J.}~\bibnamefont
  {Kong}}, \bibinfo {author} {\bibfnamefont {M.~G.}\ \bibnamefont {Chapline}},
  \ and\ \bibinfo {author} {\bibfnamefont {H.}~\bibnamefont {Dai}},\
  }\href@noop {} {\bibfield  {journal} {\bibinfo  {journal} {J. Phys. Chem. B}\
  }\textbf {\bibinfo {volume} {105}},\ \bibinfo {pages} {2890} (\bibinfo {year}
  {2001})}\BibitemShut {NoStop}%
\bibitem [{\citenamefont {Wang}\ \emph {et~al.}(2006)\citenamefont {Wang},
  \citenamefont {Liang}, \citenamefont {Liu}, \citenamefont {Wang},\ and\
  \citenamefont {Zhang}}]{Wang}%
  \BibitemOpen
  \bibfield  {author} {\bibinfo {author} {\bibfnamefont {S.}~\bibnamefont
  {Wang}}, \bibinfo {author} {\bibfnamefont {Z.}~\bibnamefont {Liang}},
  \bibinfo {author} {\bibfnamefont {T.}~\bibnamefont {Liu}}, \bibinfo {author}
  {\bibfnamefont {B.}~\bibnamefont {Wang}}, \ and\ \bibinfo {author}
  {\bibfnamefont {C.}~\bibnamefont {Zhang}},\ }\href@noop {} {\bibfield
  {journal} {\bibinfo  {journal} {Nanotechnology}\ }\textbf {\bibinfo {volume}
  {17}},\ \bibinfo {pages} {1551} (\bibinfo {year} {2006})}\BibitemShut
  {NoStop}%
\bibitem [{\citenamefont {Strano}\ \emph {et~al.}(2003)\citenamefont {Strano},
  \citenamefont {Dyke}, \citenamefont {Usrey}, \citenamefont {Barone},
  \citenamefont {Allen}, \citenamefont {Shan}, \citenamefont {Kittrell},
  \citenamefont {Hauge}, \citenamefont {Tour},\ and\ \citenamefont
  {Smalley}}]{Strano}%
  \BibitemOpen
  \bibfield  {author} {\bibinfo {author} {\bibfnamefont {M.~S.}\ \bibnamefont
  {Strano}}, \bibinfo {author} {\bibfnamefont {C.~A.}\ \bibnamefont {Dyke}},
  \bibinfo {author} {\bibfnamefont {M.~L.}\ \bibnamefont {Usrey}}, \bibinfo
  {author} {\bibfnamefont {P.~W.}\ \bibnamefont {Barone}}, \bibinfo {author}
  {\bibfnamefont {M.~J.}\ \bibnamefont {Allen}}, \bibinfo {author}
  {\bibfnamefont {H.}~\bibnamefont {Shan}}, \bibinfo {author} {\bibfnamefont
  {C.}~\bibnamefont {Kittrell}}, \bibinfo {author} {\bibfnamefont {R.~H.}\
  \bibnamefont {Hauge}}, \bibinfo {author} {\bibfnamefont {J.~M.}\ \bibnamefont
  {Tour}}, \ and\ \bibinfo {author} {\bibfnamefont {R.~E.}\ \bibnamefont
  {Smalley}},\ }\href@noop {} {\bibfield  {journal} {\bibinfo  {journal}
  {Science}\ }\textbf {\bibinfo {volume} {301}},\ \bibinfo {pages} {1519}
  (\bibinfo {year} {2003})}\BibitemShut {NoStop}%
\bibitem [{\citenamefont {Stevens}\ \emph {et~al.}(2003)\citenamefont
  {Stevens}, \citenamefont {Huang}, \citenamefont {Peng}, \citenamefont
  {Chiang}, \citenamefont {Khabashesku},\ and\ \citenamefont
  {Margrave}}]{Stevens}%
  \BibitemOpen
  \bibfield  {author} {\bibinfo {author} {\bibfnamefont {J.~L.}\ \bibnamefont
  {Stevens}}, \bibinfo {author} {\bibfnamefont {A.~Y.}\ \bibnamefont {Huang}},
  \bibinfo {author} {\bibfnamefont {H.}~\bibnamefont {Peng}}, \bibinfo {author}
  {\bibfnamefont {I.~W.}\ \bibnamefont {Chiang}}, \bibinfo {author}
  {\bibfnamefont {V.~N.}\ \bibnamefont {Khabashesku}}, \ and\ \bibinfo {author}
  {\bibfnamefont {J.~L.}\ \bibnamefont {Margrave}},\ }\href@noop {} {\bibfield
  {journal} {\bibinfo  {journal} {Nano Lett.}\ }\textbf {\bibinfo {volume}
  {3}},\ \bibinfo {pages} {331} (\bibinfo {year} {2003})}\BibitemShut {NoStop}%
\bibitem [{\citenamefont {Nakamura}\ \emph {et~al.}(2008)\citenamefont
  {Nakamura}, \citenamefont {T.~Ohana}, \citenamefont {Hasegawa},\ and\
  \citenamefont {Koga}}]{Nakamura}%
  \BibitemOpen
  \bibfield  {author} {\bibinfo {author} {\bibfnamefont {T.}~\bibnamefont
  {Nakamura}}, \bibinfo {author} {\bibfnamefont {M.~I.}\ \bibnamefont
  {T.~Ohana}}, \bibinfo {author} {\bibfnamefont {M.}~\bibnamefont {Hasegawa}},
  \ and\ \bibinfo {author} {\bibfnamefont {Y.}~\bibnamefont {Koga}},\
  }\href@noop {} {\bibfield  {journal} {\bibinfo  {journal} {Diamond and
  Reletad Materials}\ }\textbf {\bibinfo {volume} {17}},\ \bibinfo {pages}
  {559} (\bibinfo {year} {2008})}\BibitemShut {NoStop}%
\bibitem [{\citenamefont {Keren}\ \emph {et~al.}(2003)\citenamefont {Keren},
  \citenamefont {Berman}, \citenamefont {Buchstab}, \citenamefont {Sivan},\
  and\ \citenamefont {Braun}}]{Keren}%
  \BibitemOpen
  \bibfield  {author} {\bibinfo {author} {\bibfnamefont {K.}~\bibnamefont
  {Keren}}, \bibinfo {author} {\bibfnamefont {R.~S.}\ \bibnamefont {Berman}},
  \bibinfo {author} {\bibfnamefont {E.}~\bibnamefont {Buchstab}}, \bibinfo
  {author} {\bibfnamefont {U.}~\bibnamefont {Sivan}}, \ and\ \bibinfo {author}
  {\bibfnamefont {E.}~\bibnamefont {Braun}},\ }\href@noop {} {\bibfield
  {journal} {\bibinfo  {journal} {Science}\ }\textbf {\bibinfo {volume}
  {302}},\ \bibinfo {pages} {1380} (\bibinfo {year} {2003})}\BibitemShut
  {NoStop}%
\bibitem [{\citenamefont {Chelmecka}\ \emph {et~al.}(2010)\citenamefont
  {Chelmecka}, \citenamefont {Pasterny}, \citenamefont {Kupka},\ and\
  \citenamefont {Stobinski}}]{stobinski2010}%
  \BibitemOpen
  \bibfield  {author} {\bibinfo {author} {\bibfnamefont {E.}~\bibnamefont
  {Chelmecka}}, \bibinfo {author} {\bibfnamefont {K.}~\bibnamefont {Pasterny}},
  \bibinfo {author} {\bibfnamefont {T.}~\bibnamefont {Kupka}}, \ and\ \bibinfo
  {author} {\bibfnamefont {L.}~\bibnamefont {Stobinski}},\ }\href@noop {}
  {\bibfield  {journal} {\bibinfo  {journal} {J. Mol. Struct.: THEOCHEM}\
  }\textbf {\bibinfo {volume} {948}},\ \bibinfo {pages} {93} (\bibinfo {year}
  {2010})}\BibitemShut {NoStop}%
\bibitem [{\citenamefont {Doudou}\ \emph {et~al.}(2011)\citenamefont {Doudou},
  \citenamefont {Chen}, \citenamefont {Vivet}, \citenamefont {Poilane},\ and\
  \citenamefont {Ayachi}}]{Doudou}%
  \BibitemOpen
  \bibfield  {author} {\bibinfo {author} {\bibfnamefont {B.}~\bibnamefont
  {Doudou}}, \bibinfo {author} {\bibfnamefont {J.}~\bibnamefont {Chen}},
  \bibinfo {author} {\bibfnamefont {A.}~\bibnamefont {Vivet}}, \bibinfo
  {author} {\bibfnamefont {C.}~\bibnamefont {Poilane}}, \ and\ \bibinfo
  {author} {\bibfnamefont {M.}~\bibnamefont {Ayachi}},\ }\href@noop {}
  {\bibfield  {journal} {\bibinfo  {journal} {Comp. Theor. Chem.}\ }\textbf
  {\bibinfo {volume} {967}},\ \bibinfo {pages} {231} (\bibinfo {year}
  {2011})}\BibitemShut {NoStop}%
\bibitem [{\citenamefont {Zhao}\ \emph {et~al.}(2004)\citenamefont {Zhao},
  \citenamefont {Park}, \citenamefont {Han},\ and\ \citenamefont {Lu}}]{Zhao}%
  \BibitemOpen
  \bibfield  {author} {\bibinfo {author} {\bibfnamefont {J.}~\bibnamefont
  {Zhao}}, \bibinfo {author} {\bibfnamefont {H.}~\bibnamefont {Park}}, \bibinfo
  {author} {\bibfnamefont {J.}~\bibnamefont {Han}}, \ and\ \bibinfo {author}
  {\bibfnamefont {J.~P.}\ \bibnamefont {Lu}},\ }\href@noop {} {\bibfield
  {journal} {\bibinfo  {journal} {J. Phys. Chem. B}\ }\textbf {\bibinfo
  {volume} {108}},\ \bibinfo {pages} {4227} (\bibinfo {year}
  {2004})}\BibitemShut {NoStop}%
\bibitem [{\citenamefont {Zhao}\ \emph {et~al.}(2002)\citenamefont {Zhao},
  \citenamefont {Buldum}, \citenamefont {Han},\ and\ \citenamefont
  {Lu}}]{Zhao2}%
  \BibitemOpen
  \bibfield  {author} {\bibinfo {author} {\bibfnamefont {J.}~\bibnamefont
  {Zhao}}, \bibinfo {author} {\bibfnamefont {A.}~\bibnamefont {Buldum}},
  \bibinfo {author} {\bibfnamefont {J.}~\bibnamefont {Han}}, \ and\ \bibinfo
  {author} {\bibfnamefont {J.~P.}\ \bibnamefont {Lu}},\ }\href@noop {}
  {\bibfield  {journal} {\bibinfo  {journal} {Nanotechnology}\ }\textbf
  {\bibinfo {volume} {13}},\ \bibinfo {pages} {195} (\bibinfo {year}
  {2002})}\BibitemShut {NoStop}%
\bibitem [{\citenamefont {Jr.}\ and\ \citenamefont
  {Ricca}(2004)}]{Bauschlicher}%
  \BibitemOpen
  \bibfield  {author} {\bibinfo {author} {\bibfnamefont {C.~W.~B.}\
  \bibnamefont {Jr.}}\ and\ \bibinfo {author} {\bibfnamefont {A.}~\bibnamefont
  {Ricca}},\ }\href@noop {} {\bibfield  {journal} {\bibinfo  {journal} {Phys.
  Rev. B}\ }\textbf {\bibinfo {volume} {70}},\ \bibinfo {pages} {115409}
  (\bibinfo {year} {2004})}\BibitemShut {NoStop}%
\bibitem [{\citenamefont {Shirvani}\ \emph {et~al.}(2010)\citenamefont
  {Shirvani}, \citenamefont {Behesthian}, \citenamefont {Parsafar},\ and\
  \citenamefont {Hadipour}}]{Shirvani}%
  \BibitemOpen
  \bibfield  {author} {\bibinfo {author} {\bibfnamefont {B.}~\bibnamefont
  {Shirvani}}, \bibinfo {author} {\bibfnamefont {J.}~\bibnamefont
  {Behesthian}}, \bibinfo {author} {\bibfnamefont {G.}~\bibnamefont
  {Parsafar}}, \ and\ \bibinfo {author} {\bibfnamefont {N.~L.}\ \bibnamefont
  {Hadipour}},\ }\href@noop {} {\bibfield  {journal} {\bibinfo  {journal}
  {Comput. Mater. Sc.}\ }\textbf {\bibinfo {volume} {48}},\ \bibinfo {pages}
  {655} (\bibinfo {year} {2010})}\BibitemShut {NoStop}%
\bibitem [{\citenamefont {Milowska}\ \emph {et~al.}(2009)\citenamefont
  {Milowska}, \citenamefont {Birowska},\ and\ \citenamefont {Majewski}}]{appa}%
  \BibitemOpen
  \bibfield  {author} {\bibinfo {author} {\bibfnamefont {K.~Z.}\ \bibnamefont
  {Milowska}}, \bibinfo {author} {\bibfnamefont {M.}~\bibnamefont {Birowska}},
  \ and\ \bibinfo {author} {\bibfnamefont {J.~A.}\ \bibnamefont {Majewski}},\
  }\href@noop {} {\bibfield  {journal} {\bibinfo  {journal} {Acta Physica Polonica A}\ }\textbf
  {\bibinfo {volume} {116}},\ \bibinfo {pages} {841} (\bibinfo {year}
  {2009})}\BibitemShut {NoStop}%
\bibitem [{\citenamefont {Milowska}\ \emph {et~al.}(2011)\citenamefont
  {Milowska}, \citenamefont {Birowska},\ and\ \citenamefont {Majewski}}]{aip}%
  \BibitemOpen
  \bibfield  {author} {\bibinfo {author} {\bibfnamefont {K.}~\bibnamefont
  {Milowska}}, \bibinfo {author} {\bibfnamefont {M.}~\bibnamefont {Birowska}},
  \ and\ \bibinfo {author} {\bibfnamefont {J.~A.}\ \bibnamefont {Majewski}},\
  }\href@noop {} {\bibfield  {journal} {\bibinfo  {journal} {AIP Conf. Proc.}\
  }\textbf {\bibinfo {volume} {1399}},\ \bibinfo {pages} {827} (\bibinfo {year}
  {2011})}\BibitemShut {NoStop}%
\bibitem [{\citenamefont {Milowska}\ \emph {et~al.}(2012)\citenamefont
  {Milowska}, \citenamefont {Birowska},\ and\ \citenamefont
  {Majewski}}]{diamond}%
  \BibitemOpen
  \bibfield  {author} {\bibinfo {author} {\bibfnamefont {K.}~\bibnamefont
  {Milowska}}, \bibinfo {author} {\bibfnamefont {M.}~\bibnamefont {Birowska}},
  \ and\ \bibinfo {author} {\bibfnamefont {J.~A.}\ \bibnamefont {Majewski}},\
  }\href@noop {} {\bibfield  {journal} {\bibinfo  {journal} {Diamond and
  Related Materials}\ }\textbf {\bibinfo {volume} {23}},\ \bibinfo {pages}
  {167} (\bibinfo {year} {2012})}\BibitemShut {NoStop}%
\bibitem [{\citenamefont {Boys}\ and\ \citenamefont {Bernardi}(1970)}]{bsse0}%
  \BibitemOpen
  \bibfield  {author} {\bibinfo {author} {\bibfnamefont {S.~F.}\ \bibnamefont
  {Boys}}\ and\ \bibinfo {author} {\bibfnamefont {F.}~\bibnamefont
  {Bernardi}},\ }\href@noop {} {\bibfield  {journal} {\bibinfo  {journal}
  {Molecular Physics}\ }\textbf {\bibinfo {volume} {19}},\ \bibinfo {pages}
  {553} (\bibinfo {year} {1970})}\BibitemShut {NoStop}%
\bibitem [{\citenamefont {Shuttleworth}(2012)}]{bsse1}%
  \BibitemOpen
  \bibfield  {author} {\bibinfo {author} {\bibfnamefont {I.~G.}\ \bibnamefont
  {Shuttleworth}},\ }\href@noop {} {\bibfield  {journal} {\bibinfo  {journal}
  {Applied Surface Science}\ }\textbf {\bibinfo {volume} {258}},\ \bibinfo
  {pages} {7546} (\bibinfo {year} {2012})}\BibitemShut {NoStop}%
\bibitem [{\citenamefont {Hohenberg}\ and\ \citenamefont
  {Kohn}(1964)}]{Hohenberg1965}%
  \BibitemOpen
  \bibfield  {author} {\bibinfo {author} {\bibfnamefont {P.}~\bibnamefont
  {Hohenberg}}\ and\ \bibinfo {author} {\bibfnamefont {W.}~\bibnamefont
  {Kohn}},\ }\href@noop {} {\bibfield  {journal} {\bibinfo  {journal} {Phys.
  Rev.}\ }\textbf {\bibinfo {volume} {136}},\ \bibinfo {pages} {864} (\bibinfo
  {year} {1964})}\BibitemShut {NoStop}%
\bibitem [{\citenamefont {Kohn}\ and\ \citenamefont {Sham}(1965)}]{kohn}%
  \BibitemOpen
  \bibfield  {author} {\bibinfo {author} {\bibfnamefont {W.}~\bibnamefont
  {Kohn}}\ and\ \bibinfo {author} {\bibfnamefont {L.~J.}\ \bibnamefont
  {Sham}},\ }\href@noop {} {\bibfield  {journal} {\bibinfo  {journal} {Phys.
  Rev.}\ }\textbf {\bibinfo {volume} {140}},\ \bibinfo {pages} {A1133}
  (\bibinfo {year} {1965})}\BibitemShut {NoStop}%
\bibitem [{\citenamefont {Perdew}\ \emph {et~al.}(1996)\citenamefont {Perdew},
  \citenamefont {Burke},\ and\ \citenamefont {Ernzerhof}}]{Perdew1996}%
  \BibitemOpen
  \bibfield  {author} {\bibinfo {author} {\bibfnamefont {J.~P.}\ \bibnamefont
  {Perdew}}, \bibinfo {author} {\bibfnamefont {K.}~\bibnamefont {Burke}}, \
  and\ \bibinfo {author} {\bibfnamefont {M.}~\bibnamefont {Ernzerhof}},\
  }\href@noop {} {\bibfield  {journal} {\bibinfo  {journal} {Phys. Rev. Lett.}\
  }\textbf {\bibinfo {volume} {77}},\ \bibinfo {pages} {3865} (\bibinfo {year}
  {1996})}\BibitemShut {NoStop}%
\bibitem [{\citenamefont {Ordejon}\ \emph {et~al.}(1996)\citenamefont
  {Ordejon}, \citenamefont {Artacho},\ and\ \citenamefont {Soler}}]{ordejon}%
  \BibitemOpen
  \bibfield  {author} {\bibinfo {author} {\bibfnamefont {P.}~\bibnamefont
  {Ordejon}}, \bibinfo {author} {\bibfnamefont {E.}~\bibnamefont {Artacho}}, \
  and\ \bibinfo {author} {\bibfnamefont {J.~M.}\ \bibnamefont {Soler}},\
  }\href@noop {} {\bibfield  {journal} {\bibinfo  {journal} {Phys. Rev. B
  (Rapid Comm.)}\ }\textbf {\bibinfo {volume} {53}},\ \bibinfo {pages} {R10441}
  (\bibinfo {year} {1996})}\BibitemShut {NoStop}%
\bibitem [{\citenamefont {Soler}\ \emph {et~al.}(2002)\citenamefont {Soler},
  \citenamefont {Artacho}, \citenamefont {Gale}, \citenamefont {Garcia},
  \citenamefont {Junquera}, \citenamefont {Ordejon},\ and\ \citenamefont
  {Sanchez-Portal}}]{soler}%
  \BibitemOpen
  \bibfield  {author} {\bibinfo {author} {\bibfnamefont {J.~M.}\ \bibnamefont
  {Soler}}, \bibinfo {author} {\bibfnamefont {E.}~\bibnamefont {Artacho}},
  \bibinfo {author} {\bibfnamefont {J.~D.}\ \bibnamefont {Gale}}, \bibinfo
  {author} {\bibfnamefont {A.}~\bibnamefont {Garcia}}, \bibinfo {author}
  {\bibfnamefont {J.}~\bibnamefont {Junquera}}, \bibinfo {author}
  {\bibfnamefont {P.}~\bibnamefont {Ordejon}}, \ and\ \bibinfo {author}
  {\bibfnamefont {D.}~\bibnamefont {Sanchez-Portal}},\ }\href@noop {}
  {\bibfield  {journal} {\bibinfo  {journal} {J. Phys.: Condens. Matter}\
  }\textbf {\bibinfo {volume} {14}},\ \bibinfo {pages} {2745} (\bibinfo {year}
  {2002})}\BibitemShut {NoStop}%
\bibitem [{Note1()}]{Note1}%
   \bibinfo {note} {Similar observation was marked in Ref.\protect
  \rev@citealpnum {Rosi2007}.}\BibitemShut {NoStop}%
\bibitem [{\citenamefont {Liu}\ \emph {et~al.}(1999)\citenamefont {Liu},
  \citenamefont {Casavant}, \citenamefont {Cox}, \citenamefont {Walters},
  \citenamefont {Boul}, \citenamefont {Lu}, \citenamefont {Rimberg},
  \citenamefont {Smith}, \citenamefont {Colbert},\ and\ \citenamefont
  {Smalley}}]{Liu99}%
  \BibitemOpen
  \bibfield  {author} {\bibinfo {author} {\bibfnamefont {J.}~\bibnamefont
  {Liu}}, \bibinfo {author} {\bibfnamefont {M.~J.}\ \bibnamefont {Casavant}},
  \bibinfo {author} {\bibfnamefont {M.}~\bibnamefont {Cox}}, \bibinfo {author}
  {\bibfnamefont {D.}~\bibnamefont {Walters}}, \bibinfo {author} {\bibfnamefont
  {P.}~\bibnamefont {Boul}}, \bibinfo {author} {\bibfnamefont {W.}~\bibnamefont
  {Lu}}, \bibinfo {author} {\bibfnamefont {A.}~\bibnamefont {Rimberg}},
  \bibinfo {author} {\bibfnamefont {K.}~\bibnamefont {Smith}}, \bibinfo
  {author} {\bibfnamefont {D.~T.}\ \bibnamefont {Colbert}}, \ and\ \bibinfo
  {author} {\bibfnamefont {R.~E.}\ \bibnamefont {Smalley}},\ }\href@noop {}
  {\bibfield  {journal} {\bibinfo  {journal} {Chem. Phys. Lett.}\ }\textbf
  {\bibinfo {volume} {303}},\ \bibinfo {pages} {125} (\bibinfo {year}
  {1999})}\BibitemShut {NoStop}%
\bibitem [{\citenamefont {Owens}(2007)}]{Owens}%
  \BibitemOpen
  \bibfield  {author} {\bibinfo {author} {\bibfnamefont {F.}~\bibnamefont
  {Owens}},\ }\href@noop {} {\bibfield  {journal} {\bibinfo  {journal}
  {Materials Letters}\ }\textbf {\bibinfo {volume} {61}},\ \bibinfo {pages}
  {1997} (\bibinfo {year} {2007})}\BibitemShut {NoStop}%
\bibitem [{Note2()}]{Note2}%
  \bibinfo {note} {Rosi used finite (relatively small) CNTs and hybrid B3LYP
  functional.}\BibitemShut {NoStop}%
\bibitem [{\citenamefont {Matsuda}\ \emph {et~al.}(2010)\citenamefont
  {Matsuda}, \citenamefont {Tahir-Kheli},\ and\ \citenamefont {III}}]{Matsuda}%
  \BibitemOpen
  \bibfield  {author} {\bibinfo {author} {\bibfnamefont {Y.}~\bibnamefont
  {Matsuda}}, \bibinfo {author} {\bibfnamefont {J.}~\bibnamefont
  {Tahir-Kheli}}, \ and\ \bibinfo {author} {\bibfnamefont {W.~A.~G.}\
  \bibnamefont {III}},\ }\href@noop {} {\bibfield  {journal} {\bibinfo
  {journal} {J. Phys. Chem. Lett.}\ }\textbf {\bibinfo {volume} {1}},\ \bibinfo
  {pages} {2946} (\bibinfo {year} {2010})}\BibitemShut {NoStop}%
\bibitem [{\citenamefont {Cao}\ \emph {et~al.}(2001)\citenamefont {Cao},
  \citenamefont {Yan}, \citenamefont {Ding},\ and\ \citenamefont {Wang}}]{Cao}%
  \BibitemOpen
  \bibfield  {author} {\bibinfo {author} {\bibfnamefont {J.~X.}\ \bibnamefont
  {Cao}}, \bibinfo {author} {\bibfnamefont {X.~H.}\ \bibnamefont {Yan}},
  \bibinfo {author} {\bibfnamefont {J.~W.}\ \bibnamefont {Ding}}, \ and\
  \bibinfo {author} {\bibfnamefont {D.~L.}\ \bibnamefont {Wang}},\ }\href@noop
  {} {\bibfield  {journal} {\bibinfo  {journal} {J. Phys.:Condens. Matter}\
  }\textbf {\bibinfo {volume} {13}},\ \bibinfo {pages} {L271} (\bibinfo {year}
  {2001})}\BibitemShut {NoStop}%
\bibitem [{\citenamefont {Kienle}\ \emph {et~al.}(2006)\citenamefont {Kienle},
  \citenamefont {Cerda},\ and\ \citenamefont {Ghosh}}]{Kienle}%
  \BibitemOpen
  \bibfield  {author} {\bibinfo {author} {\bibfnamefont {D.}~\bibnamefont
  {Kienle}}, \bibinfo {author} {\bibfnamefont {J.~I.}\ \bibnamefont {Cerda}}, \
  and\ \bibinfo {author} {\bibfnamefont {A.~W.}\ \bibnamefont {Ghosh}},\
  }\href@noop {} {\bibfield  {journal} {\bibinfo  {journal} {J. Appl. Phys.}\
  }\textbf {\bibinfo {volume} {100}},\ \bibinfo {pages} {043714} (\bibinfo
  {year} {2006})}\BibitemShut {NoStop}%
\bibitem [{\citenamefont {Gulseren}\ \emph {et~al.}(2002)\citenamefont
  {Gulseren}, \citenamefont {Yildirim}, \citenamefont {Ciraci},\ and\
  \citenamefont {Kilic}}]{Gulseren}%
  \BibitemOpen
  \bibfield  {author} {\bibinfo {author} {\bibfnamefont {O.}~\bibnamefont
  {Gulseren}}, \bibinfo {author} {\bibfnamefont {T.}~\bibnamefont {Yildirim}},
  \bibinfo {author} {\bibfnamefont {S.}~\bibnamefont {Ciraci}}, \ and\ \bibinfo
  {author} {\bibfnamefont {C.}~\bibnamefont {Kilic}},\ }\href@noop {}
  {\bibfield  {journal} {\bibinfo  {journal} {Phys. Rev. B}\ }\textbf {\bibinfo
  {volume} {65}},\ \bibinfo {pages} {155410} (\bibinfo {year}
  {2002})}\BibitemShut {NoStop}%
\bibitem [{\citenamefont {Silva}\ \emph {et~al.}(2004)\citenamefont {Silva},
  \citenamefont {Fagan},\ and\ \citenamefont {Mota}}]{Silva}%
  \BibitemOpen
  \bibfield  {author} {\bibinfo {author} {\bibfnamefont {L.~B.}\ \bibnamefont
  {Silva}}, \bibinfo {author} {\bibfnamefont {S.~B.}\ \bibnamefont {Fagan}}, \
  and\ \bibinfo {author} {\bibfnamefont {R.}~\bibnamefont {Mota}},\ }\href@noop
  {} {\bibfield  {journal} {\bibinfo  {journal} {Nano Lett.}\ }\textbf
  {\bibinfo {volume} {4}},\ \bibinfo {pages} {65} (\bibinfo {year}
  {2004})}\BibitemShut {NoStop}%
\bibitem [{\citenamefont {Thostenson}\ \emph {et~al.}(2001)\citenamefont
  {Thostenson}, \citenamefont {Ren},\ and\ \citenamefont
  {Chou}}]{Thostenson2001}%
  \BibitemOpen
  \bibfield  {author} {\bibinfo {author} {\bibfnamefont {E.~T.}\ \bibnamefont
  {Thostenson}}, \bibinfo {author} {\bibfnamefont {Z.}~\bibnamefont {Ren}}, \
  and\ \bibinfo {author} {\bibfnamefont {T.-W.}\ \bibnamefont {Chou}},\
  }\href@noop {} {\bibfield  {journal} {\bibinfo  {journal} {Composites Science
  and Technology}\ }\textbf {\bibinfo {volume} {61}},\ \bibinfo {pages} {1899}
  (\bibinfo {year} {2001})}\BibitemShut {NoStop}%
\end{thebibliography}
\end{document}